\documentclass[preprint,showpacs,preprintnumbers,amsmath,amssymb]{revtex4}

\usepackage{graphicx}
\usepackage{dcolumn}
\usepackage{bm}
\usepackage{multirow}

\def\AB#1#2#3{\langle#1|#2|#3]}

\def\fl#1{{#1}^\flat}

\def\e{\epsilon}
\def\ra{\rangle}
\def\la{\langle}
\def\cg{c_\Gamma}
\def\rG{r_{\Gamma}}
\newcommand{\slsh}{\rlap{$\;\!\!\not$}}     
\def\spa#1.#2{\left\langle#1\,#2\right\rangle}
\def\spb#1.#2{\left[#1\,#2\right]}
\def\spaa#1.#2.#3{\langle\mskip-1mu{#1} 
                  | #2 | {#3}\mskip-1mu\rangle}
\def\spbb#1.#2.#3{[\mskip-1mu{#1}
                  | #2 | {#3}\mskip-1mu]}
\def\spbab#1.#2.#3.#4{[\mskip-1mu{#1} 
                  | #2 | #3 | {#4}\mskip-1mu]}
\def\spaba#1.#2.#3.#4{\langle\mskip-1mu{#1} 
                  | #2 | #3 | {#4}\mskip-1mu\rangle}
\def\spab#1.#2.#3{\langle\mskip-1mu{#1} 
                  | #2 | {#3}\mskip-1mu]}
\def\spba#1.#2.#3{[\mskip-1mu{#1} 
                  | #2 | {#3}\mskip-1mu\rangle}
\newcommand{\ta}[1]{#1\hspace{-.42em}/\hspace{-.07em}}

\def\ellb{{\overline{\ell}}}
\def\beqn{\begin{eqnarray}}
\def\eeqn{\end{eqnarray}}
\def\x#1{\langle#1\rangle}

\def\nn{\nonumber}
\def\lc{{\rm lc}}
\def\sl{{\rm sl}}
\def\cb{{\rm cb}}
\def\lf{{\rm lf}}
\def\hf{{\rm hf}}
\def\cA#1{{\cal A}_{#1}}
\def\beq{\begin{equation}}
\def\eeq{\end{equation}}
\def\q{{\vphantom{\qb}{q}}}
\def\l{{\vphantom{\lb}{l}}}
\def\Q{{\vphantom{\Qb}{Q}}}
\def\qb{{\bar q}}
\def\lb{{\bar l}}
\def\Qb{{\bar Q}}
\def\prop#1{{\cal P}_{#1}}
\def\Atree{A^{\rm tree}}
\def\Alc{A^\lc}
\def\Asl{A^\sl}
\def\Acb{A^\cb}
\def\Alf{A^\lf}
\def\Ahf{A^\hf}
\def\cAtree{{\cal A}^{\rm tree}}
\def\jb{{\; \bar{j}}}

\def\li{{\rm Li_2}}
\def\braa#1{\langle#1|}
\def\keta#1{|#1\rangle}
\def\brab#1{[#1|}
\def\ketb#1{|#1]}

\begin{document}
\preprint{FERMILAB-Pub-10-482-T}
\title{QCD corrections to the hadronic production of a heavy quark pair and a $W$-boson including decay correlations}
\author{Simon Badger}
\email{simon.badger@nbi.dk}
\affiliation{Niels Bohr International Academy and Discovery Center,
The Niels Bohr Institute, Blegdamsvej 17, DK-2100 Copenhagen, Denmark}
\author{John M. Campbell}
\email{johnmc@fnal.gov}
\author{R. K. Ellis}%
\email{ellis@fnal.gov}
\affiliation{
Theoretical Physics Department, Fermi National Accelerator Laboratory,\\
P.~O.~Box 500, Batavia, IL 60510,  USA}

\date{\today}

\begin{abstract}
We perform an analytic calculation of the one-loop amplitude for the $W$-boson mediated 
process $0 \to d \bar{u} Q \bar{Q} \bar{\ell} \ell$ retaining the mass for the quark $Q$. 
The momentum of each of the massive quarks 
is expressed as the sum of two massless momenta and the corresponding heavy quark spinor 
is expressed as a sum of two massless spinors.
Using a special choice for the heavy quark spinors we obtain analytic
expressions for the one-loop amplitudes which are amenable to 
fast numerical evaluation.
The full next-to-leading order (NLO) calculation of ${\rm hadron}~+~{\rm hadron} \to W(\to e \nu) b \bar{b}$
with massive $b$-quarks is included in the program MCFM. 
A comparison is performed with previous published work.
\end{abstract}

\maketitle

\section{Introduction}
\begin{figure}
\begin{center}
\includegraphics[angle=270,scale=0.5]{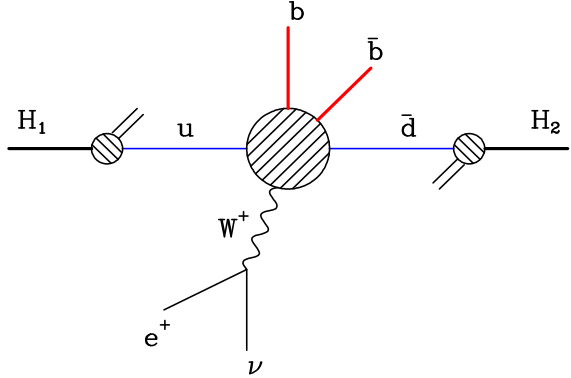}
\end{center}
\caption{Example of a partonic process contributing to $W b \bar{b}$ production}
\label{ppbarWbb}
\end{figure}

One of the most interesting channels currently under study at the Tevatron
and the LHC is the final state containing a $W$-boson and jets, where some 
or all of the produced jets are tagged as containing a bottom quark. Several 
interesting partonic processes contribute to this final state, for example,
\begin{itemize}
\item 
$W+H \, (\to b \bar{b})$
\item 
$W+Z \, (\to b \bar{b})$
\item 
$\bar{b}+t \, (\to W^+ +b)$
\item 
$\bar{t}+t \, (\to W^+ +b)$
\end{itemize}
A background to these processes involving heavy bosons and fermions,
is the QCD and electroweak process occurring in the collision of hadrons $H_1$ and $H_2$,
\beq
H_1+H_2 \to W + b + \bar{b} +X \; . 
\eeq
An example of a partonic subprocess contributing to this process is shown in Fig.~\ref{ppbarWbb}. 
Next-to-leading corrections to this process 
were first considered in \cite{Ellis:1998fv},
working in the approximation 
in which the $b$-quark is considered massless. Since in many analyses 
the $b$-quark is required to have a minimum $p_T$, typically $15$~GeV or more, in
order to be efficiently tagged, the neglect of the mass of the bottom quark
is expected to be a good approximation. 
 
After these initial studies, this same process was considered in
refs.~\cite{Febres Cordero:2006sj,Cordero:2009kv,Cordero:2010qn,Campbell:2008hh},
without making the approximation $m_b=0$. These studies, performed
without the inclusion of decay products of the $W$ boson effectively
confirmed that the $m_b=0$ approximation is good to a few percent,
with the difference parametrically suppressed as $m_b^2/p_T^2$.
Thus retaining the mass of the $b$-quark extends the prediction to lower values of 
$p_T$. In addition, it allows us treat the
case where the two $b$-quarks end up in the same jet and the case when one
$b$-quark is either too soft or too forward to be tagged. These
kinematic configurations can be important for the Higgs search at the Tevatron and
the LHC~\cite{:2010ar,Butterworth:2008iy}.

The purpose of this paper is to repeat the calculation of
ref.~\cite{Febres Cordero:2006sj}, i.e.\ including the effects of a
finite $b$-quark mass, but also including the spin correlations present in the
$W$-boson decay. The calculation is performed using the spinor
helicity formalism, with the calculated amplitudes represented as
analytic formulae. Although the formulae are not compact enough to be
presented in their entirety here, they do lead to an efficient and
numerically stable code.

Following a four-dimensional unitarity based
approach~\cite{Bern:1994zx,Bern:1994cg}, we will construct the logarithmic parts of
the virtual amplitude using multiple cuts. 
Developments to this technique utilizing complex momenta~\cite{Britto:2004nc,Britto:2005ha,Britto:2006sj,Forde:2007mi}
allow us to compute analytic expressions for
the coefficients of the known scalar integrals. 

The bulk of our paper is dedicated to a description of the calculation 
of the one-loop corrections to the process 
$0 \to d \bar{u} Q \bar{Q} \bar{\ell} \ell$ retaining the mass for 
the quark $Q$.  The detailed plan of this paper is as follows. In section II 
we present our method for dealing with massive spinors using spinor helicity
techniques. Section III gives a precise definition of the amplitude that we 
wish to calculate, including the decomposition into colour stripped amplitudes
and the further decomposition into one-loop primitive amplitudes. Sections IV, V and VI
illustrate our calculation of the leading colour primitive amplitude, $A^{lc}_6$,
of the sub-leading colour primitive, $\Asl_6$, and of the primitive amplitudes containing
a closed loop of fermions, $\Alf_6$ and $\Ahf_6$. 
Section VII presents the renormalization counterterms. After describing the implementation 
of the calculation into MCFM in section VIII and comparing with earlier work, we draw some conclusions in 
section IX.

\section{Treatment of massive spinors}
A method for dealing with a massive particle in the context 
of the spinor helicity method has been given by 
Kleiss and Stirling~\cite{Kleiss:1985yh}.
A massive momentum can always be represented as a sum of two massless momenta.
Thus the spinor solution for a massive particle can be expressed in terms 
of massless spinors by decomposing the physical momentum in terms of the 
two massless momenta. If we are ultimately 
going to sum over the spin degrees of freedom, 
the only constraint that the massive spinors must satisfy is that they 
should give the standard result for the spin sum
after averaging over polarizations, namely,
\begin{eqnarray}
\sum_{s=\pm} u_s(p,m) \bar{u}_s(p,m) = \slsh{p} +m \;, \nonumber \\
\sum_{s=\pm} v_s(p,m) \bar{v}_s(p,m) = \slsh{p} -m \;.
\end{eqnarray}

We can decompose a massive vector into two massless vectors by
introducing an arbitrary massless reference vector, $\eta$,
\begin{equation}
        p = \fl p + \frac{m_p^2}{\spab{\eta}.{p}.{\eta}} \eta.
        \label{eq:massdecomp}
\end{equation}
In this equation $\fl{p}$ is a massless vector.  
The details of our spinor product notation are given in Appendix~\ref{Notation}.
The definitions of massive external spinor wave functions are,
\begin{eqnarray} \label{massivespinorwitheta}
        \bar{u}_\pm(p,m;{\fl p},\eta) &= \frac{1}{\spa{\eta\mp|}.{\fl p\pm }}{\la\eta\mp|}(\slsh{p}+m) \;, \nn \\
        v_\pm(p,m;{\fl p},\eta) &= \frac{1}{\spa{\fl p\mp |}.{\eta\pm}} {(\slsh{p}-m)|\eta\pm\ra} \;,
\end{eqnarray}
where the subscripts $\pm$ label the spin degrees of freedom. In the massless limit these labels 
correspond to the helicity quantum numbers, but in the massive case they have no such interpretation.
Treating the spinors as independent functions of $\fl{p}$ and $\eta$ i.e.\ ignoring the constraint in 
Eq.~(\ref{eq:massdecomp}) we can show 
using simple manipulations 
that,
\begin{eqnarray}
\frac{\spb{\fl p} . {\eta}}{m} \; \bar{u}_{-}(p,m;\eta,{\fl p}) &=& \bar{u}_{+}(p,m;{\fl p}, \eta) \;,\nn \\
\frac{\spa{\fl p} . {\eta}}{m} \; v_{+}(p,m;\eta,{\fl p}) &=& v_{-}(p,m;{\fl p}, \eta) \;.
\end{eqnarray}
This has the attractive feature that amplitudes with different spin labels
can be obtained from one another by exchanging $\fl{p}$ and $\eta$. This method has been used
in the calculation of one-loop corrections to top production~\cite{Badger:2010zz,Badger:2010wf}.

\subsection{Special choice for massive spinors}
In this paper, however, we adopt a different approach.  By making a
specific choice for the vector $\eta$ in terms of other vectors in the
problem we can simplify the calculation of individual amplitudes. In
addition, we shall find that for our particular choice of the vector
$\eta$, one-loop results for the colour suppressed primitive amplitude can be
obtained directly from the corresponding massless amplitude.

The implementation of the Kleiss-Stirling 
scheme appropriate for the case where we have pairs of massive particles 
is due to Rodrigo~\cite{Rodrigo:2005eu}. The two massive momenta
$p_2$ and $p_3$ with equal masses, corresponding to the 
momenta of the antiquark and quark
respectively, are written in terms of lightlike momenta $k_2$ and $k_3$,
\begin{eqnarray} \label{Rodrigodecomposition}
p_2^\mu &=& 
\frac{1+\beta}{2} \, k_2^\mu + \frac{1-\beta}{2} \, k_3^\mu \;, \nn \\
p_3^\mu &=& 
\frac{1+\beta}{2} \, k_3^\mu + \frac{1-\beta}{2} \, k_2^\mu \;, 
\end{eqnarray}
where,
\beq \label{betadef}
\beta=\sqrt{1-4m^2/s_{23}} \;, \qquad \beta_{\pm}=\frac{1}{2}(1\pm \beta) \;,
\eeq
and  $s_{23}=(p_2+p_3)^2 \equiv 2 k_2 \cdot k_3$. 
The decomposition of Eq.~(\ref{Rodrigodecomposition}) has the advantage that
momentum conservation is preserved, $p_2 + p_3 = k_2+k_3$.
The inverse transformation is given by,
\begin{eqnarray} \label{inversetransformation}
k_2^\mu &=&  
\frac{1+\beta}{2\beta} \, p_2^\mu - \frac{1-\beta}{2\beta} \, p_3^\mu~ \;, \nn \\
k_3^\mu &=& 
\frac{1+\beta}{2\beta} \, p_3^\mu - \frac{1-\beta}{2\beta} \, p_2^\mu \;.
\end{eqnarray}
In the rest of this paper we shall 
denote massive vectors by $p_i,\; (p_i^2\neq 0)$ and massless vectors by $k_i,\; (k_i^2=0)$. For the 
massless vectors $k_i$ we shall further define massless spinors,
\beqn
u_{-}(k_i)= |i-\rangle &= |i] \;, \qquad u_{+}(k_i)&= |i+\rangle = |i\rangle \;, \nn \\
\bar{u}_{-}(k_i)= \langle i-| &= \langle i| \;, \qquad \bar{u}_{+}(k_i)&= [i+| = [i| \;. 
\eeqn
In terms of the two massless momenta in Eq.~(\ref{inversetransformation}) 
the explicit results for the massive solutions 
of the Dirac equation are,
\begin{equation}
\label{Massivespinordefs}
\bar{u}_\pm(p_3,m) = \frac{\beta_+^{-1/2}}{\x{2^\mp | 3^\pm}} 
\langle 2^\mp | \, (\ta{p}_3+m) \;, \qquad
v_\pm(p_2,m) = \frac{\beta_+^{-1/2}}{\x{2^\mp | 3^\pm}}
(\ta{p}_2-m) \, |3^\pm \rangle \;.
\end{equation}
This corresponds to choosing,
\beqn
\eta_2 &= k_3, \;\; p_2^\flat &= \beta_+ k_2 \;,\nn \\
\eta_3 &= k_2, \;\; p_3^\flat &= \beta_+ k_3\;,
\eeqn
in Eqs.~(\ref{eq:massdecomp},\ref{massivespinorwitheta}).

With this choice, the results for the massive quark current with
spin labels $\{+,-\}$ and $\{-,+\}$ have the same form
as they would have in the massless
limit (i.e.\ in the limit in which $p_i \to k_i$ for $i=2,3$),
\beq
\label{helicityconserving}
S^\mu(3^\pm_\Q,2^\mp_\Qb) = \bar{u}_\pm(p_3,m) \gamma^\mu v_\mp (p_2,m) 
=\bar{u}_\pm(k_3) \gamma^\mu v_\mp (k_2) 
\equiv  \x{3^\pm|\gamma^\mu| 2^\pm}~.
\eeq
The results for spin labels $\{-,-\}$ and $\{+,+\}$ on the heavy quark current are,
\beqn
\label{helicityviolating}
S^\mu(3^-_\Q,2^-_\Qb) &=& \bar{u}_-(p_3,m) \gamma^\mu v_- (p_2,m) 
= 2 {\cal N}_{--} \; (k_2-k_3)^\mu \;,
 \nonumber \\
S^\mu(3^+_\Q,2^+_\Qb) &=& \bar{u}_+(p_3,m) \gamma^\mu v_+ (p_2,m) 
= 2 {\cal N}_{++} \; (k_2-k_3)^\mu \;,
\eeqn
where the overall normalization is given by,
\begin{equation} \label{normalization}
{\cal N}_{--}= \frac{m} {\spb2.3}, \;\;\;
{\cal N}_{++} = \frac{m} {\spa2.3} \;.
\end{equation}
Contracting these equations with a Dirac matrix we obtain,
\beqn
\label{rightleft}
S^\mu(3^+_\Q,2^-_\Qb) \otimes \left\{\gamma_\mu\right\} &=& 
2 \left\{ |2\rangle [3|+ |3] \langle 2| \right\} \;, \\
\label{leftright}
S^\mu(3^-_\Q,2^+_\Qb) \otimes \left\{\gamma_\mu\right\} &=& 
2 \left\{ |3\rangle [2|+ |2] \langle 3| \right\} \;, \\
\label{leftleft}
S^\mu(3^-_\Q,2^-_\Qb) \otimes \left\{\gamma_\mu\right\} &=& 
2 {\cal N}_{--} \left\{ |2\rangle [2|-|3\rangle [3|+ |2] \langle 2| - |3] \langle 3| \right\} \;, \\
\label{rightright}
S^\mu(3^+_\Q,2^+_\Qb) \otimes \left\{\gamma_\mu\right\} &=& 
2 {\cal N}_{++} \left\{ |2\rangle [2|-|3\rangle [3|+ |2] \langle 2| - |3] \langle 3| \right\} \;.
\eeqn
As an example of the use of these spinors, in Appendix~\ref{gQbQg} we outline the calculation of
the tree-level amplitudes for $g\Qb\Q g$ scattering, which will appear later as an ingredient
in the calculation of our one-loop amplitudes.

\section{Setup}
\begin{figure}
\begin{center}
\includegraphics[angle=270,scale=0.5]{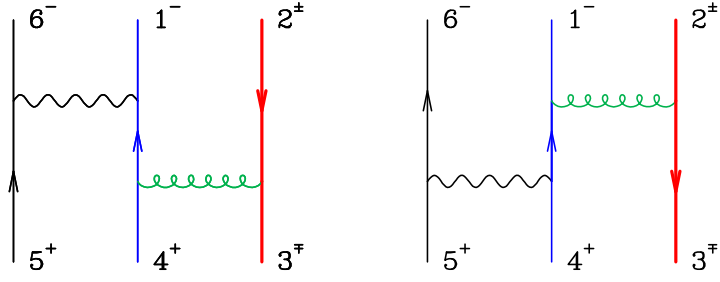}
\end{center}
\caption{Feynman graphs that enter the calculation of the lowest order amplitude.
The massive quarks are represented by the heavy (red) line, the wavy line denotes a $W$ boson
and as usual the helical line denotes a gluon.}
\label{lowestorder}
\end{figure}
We shall consider the process,
\beq
0 \to q(k_1)+\Qb(p_2)+\Q(p_3)+\qb(k_4) + \bar{\ell}(k_5) + \ell(k_6) \;,
\eeq
both at tree level and including the one-loop QCD corrections.
The process is mediated by the exchange of a $W$-boson which decays
into an antilepton-lepton pair with momenta labelled by $k_5$ and
$k_6$, as shown in Fig.~\ref{lowestorder}. The V-A structure of the charged weak interaction
ensures that the lepton and massless-quark lines will have fixed helicity.  Thus
the outgoing lepton (6) will be always left-handed and the outgoing
antilepton (5) always right-handed. With this understanding we
can often drop the lepton labels from the specification of the
amplitude.

\subsection{Colour decomposition}
As noted above we shall suppress the labels for the 
outgoing leptons $5$ and $6$ which play no role in the colour decomposition.
The colour decomposition for the tree graphs is,
\beq
\cAtree_6 (1_\q, 2_\Qb, 3_\Q, 4_\qb)  = 
  g_W^2 g^2 \; \prop{W}(s_{56}) \; \Atree_6(1_\q,2_\Qb,3_\Q,4_\qb) 
 \; \Bigl(\delta_{j_1}^{\jb_2} \,  \delta_{j_3}^{\jb_4}  \, 
 -\frac{1}{N_c} \delta_{j_1}^{\jb_4}\,  \delta_{j_3}^{\jb_2} \,\Bigr)\; .
\label{TreeColourDecomposition}
\eeq
In Eq.~(\ref{TreeColourDecomposition}) we have introduced the strong coupling constant, $g$, 
and the weak coupling, $g_W$, defined through,
\beq
\alpha_S = \frac{g^2}{4 \pi} \;, \qquad \frac{G_F}{\sqrt{2}} = \frac{g_W^2}{8 M_W^2} \;.
\eeq
The Breit-Wigner factor $\prop{W}(s)$ is given by, 
\beq
\prop{W}(s) = \frac{s}{s - M_W^2 + i \,\Gamma_W \, M_W}\,.
\eeq
The indices $j_1,j_3,(\jb_2,\jb_4)$ denote the colour labels of the corresponding quark (antiquark) lines.

At one loop the colour decomposition is given by,
\begin{eqnarray}
&&\cA{6}^{1\rm -loop} (1_\q ,2_\Qb,3_\Q,4_\qb)  =  
g_W^2 g^4 \cg  \; \prop{W}(s_{56}) \; 
\nn \\
& \times &
\Bigl[ N_c \,  \delta_{j_1}^{\jb_2} \, \delta_{j_3}^{\jb_4}  \, 
         A_{6;1}(1_\q,2_\Qb,3_\Q,4_\qb)
   + \delta_{j_1}^{\jb_4}\,  \delta_{j_3}^{\jb_2} \,
           A_{6;2} (1_\q,2_\Qb,3_\Q,4_\qb) \Bigr] \; ,
\label{qqqqDecomp}
\end{eqnarray}
where the overall factor $\cg$ is,
\begin{equation}
\cg = \frac{1}{(4\pi)^{2-\e}}
\frac{\Gamma(1+\e)\Gamma^2(1-\e)}{\Gamma(1-2\e)}\ .
\end{equation}
The interference of the one-loop amplitude with the lowest order, summed over initial and final 
colours, is given by,
\begin{eqnarray}
\sum_{{\rm colours}} [\cA{6}^* \cA{6}]_{{\rm NLO}} 
  & = &  2 g_W^4 \, g^6 \cg \, (N_c^2-1) N_c \; |\prop{W}(s_{56})|^2 
\nonumber \\
& \times &
{\rm Re} \Big\{
   [A_6^{\rm tree}(1_q,2,3,4_\qb)]^*  A_{6;1}(1_q,2,3,4_\qb) \Big\}\; .
\label{looptreeint}
\end{eqnarray}
Therefore $A_{6;2}$ plays no role in the calculation of NLO corrections.

\subsection{Tree level amplitudes}
The tree level amplitudes can be calculated using the diagrams of Fig.~\ref{lowestorder}.
The $\{ -+ \}$ amplitude, expressed in terms of the momenta $k_i$, is identical 
to the massless result as a consequence of Eq.~(\ref{helicityconserving}),
\beqn \label{Treehelicityconserving}
 - i A^{{\rm tree}}_6(1_\q^-,2_\Qb^+,3_\Q^-,4_\qb^+,5_\ellb^+,6_\ell^-) &=&
  \left[ \frac{\spa1.3\spb4.5\spab6.{(1+3)}.2}{s_{23} s_{56} s_{123}}
-        \frac{\spb4.2\spa1.6\spba5.{(2+4)}.3}{s_{23} s_{56} s_{234}} 
\right] \nn \\
&\equiv&  \left[ \Bigg\{ \frac{\spa1.3\spb4.5\spab6.{(1+3)}.2}{s_{23} s_{56} s_{123}} \Bigg\}
 -\Bigg\{{\rm flip}\Bigg\} \right] \;,
\eeqn
where we have introduced the symmetry,
\beq
{\rm flip} : 
(1 \leftrightarrow 4),\; 
(2 \leftrightarrow 3),\; 
(5 \leftrightarrow 6),\;  [ ~ ] \leftrightarrow \langle ~ \rangle \; .
\label{flip1def}
\eeq
The $\{ - - \}$  amplitude is given by,
\beqn \label{Treehelicityviolating}
&-&i A^{{\rm tree}}_6(1_\q^-, 2_\Qb^-,3_\Q^-, 4_\qb^+,5_\ellb^+,6_\ell^-) = {\cal N}_{--}  \nn \\
&&\hskip 1.5cm \times \left[ 
 \Bigg\{\spb4.5  \left( 
\frac{\spab6.{(1+2)}.3 \spa3.1 -\spab6.{(1+3)}.2 \spa2.1}
 {s_{23} s_{56} s_{123}}\right)\Bigg\}
+\Bigg\{{\rm flip}\Bigg\} \right] \;,
\eeqn
with the mass-dependent normalization factor given in Eq.~(\ref{normalization}).
The amplitudes for 
$A^{{\rm tree}}_6(1^-,2^-,3^+,4^+,6^-,5^+)$
and $A^{{\rm tree}}_6(1^-,2^+,3^+,4^+,6^-,5^+)$
can be easily obtained by reference to 
Eqs.~(\ref{helicityconserving},\ref{helicityviolating}),
\beqn
A^{{\rm tree}}_6(1_\q^-,2_\Qb^-, 3_\Q^+,4_\qb^+,5_\ellb^+,6_\ell^-)
 &=& A^{{\rm tree}}_6(1_\q^-,3_\Qb^+, 2_\Q^-,4_\qb^+,5_\ellb^+,6_\ell^-) \;, \\
A^{{\rm tree}}_6(1_\q^-,2_\Qb^+, 3_\Q^+,4_\qb^+,5_\ellb^+,6_\ell^-)
 &=& -{\rm flip} \left[A^{{\rm tree}}_6(1_\q^-,2_\Qb^-, 3_\Q^-,4_\qb^+,5_\ellb^+,6_\ell^-) \right] \;.
\label{pp_from_mm}
\eeqn

\subsection{Decomposition of one-loop amplitudes into primitive amplitudes}
\begin{figure}
\begin{center}
\includegraphics[angle=270,scale=0.5]{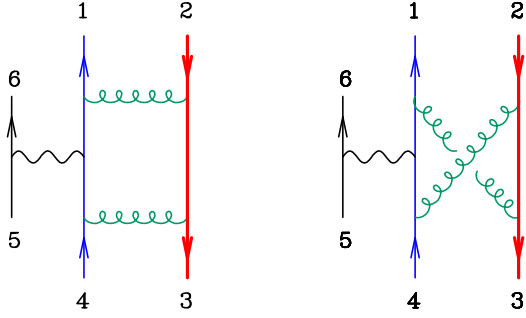}
\end{center}
\caption{Parent diagrams for the leading colour primitive $\Alc_6(1,2,3,4)$ and crossed box primitive $\Acb_6(1,2,3,4)$.}
\label{ABLC}
\end{figure}

\begin{figure}
\begin{center}
\includegraphics[angle=270,scale=0.5]{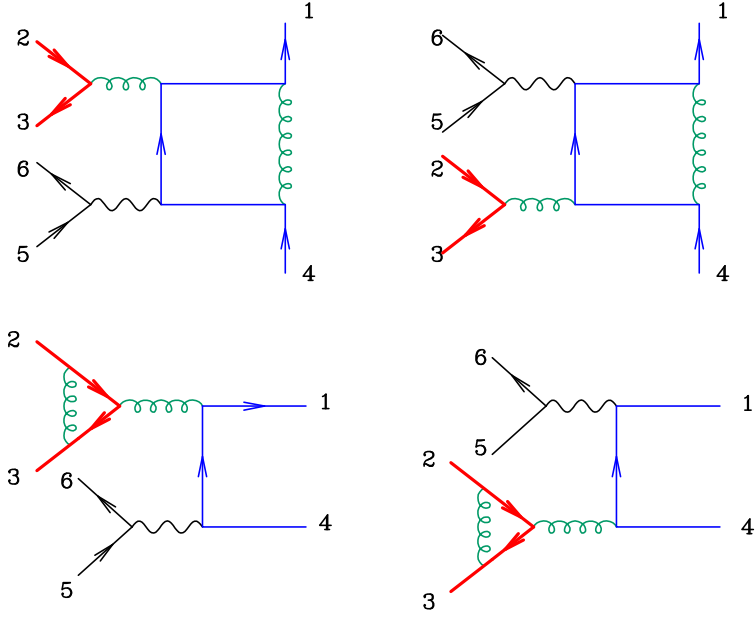}
\end{center}
\caption{Parent diagrams for the subleading colour primitive amplitude, $\Asl_6(1,2,3,4)$.}
\label{Asl}
\end{figure}

\begin{figure}
\begin{center}
\includegraphics[angle=270,scale=0.45]{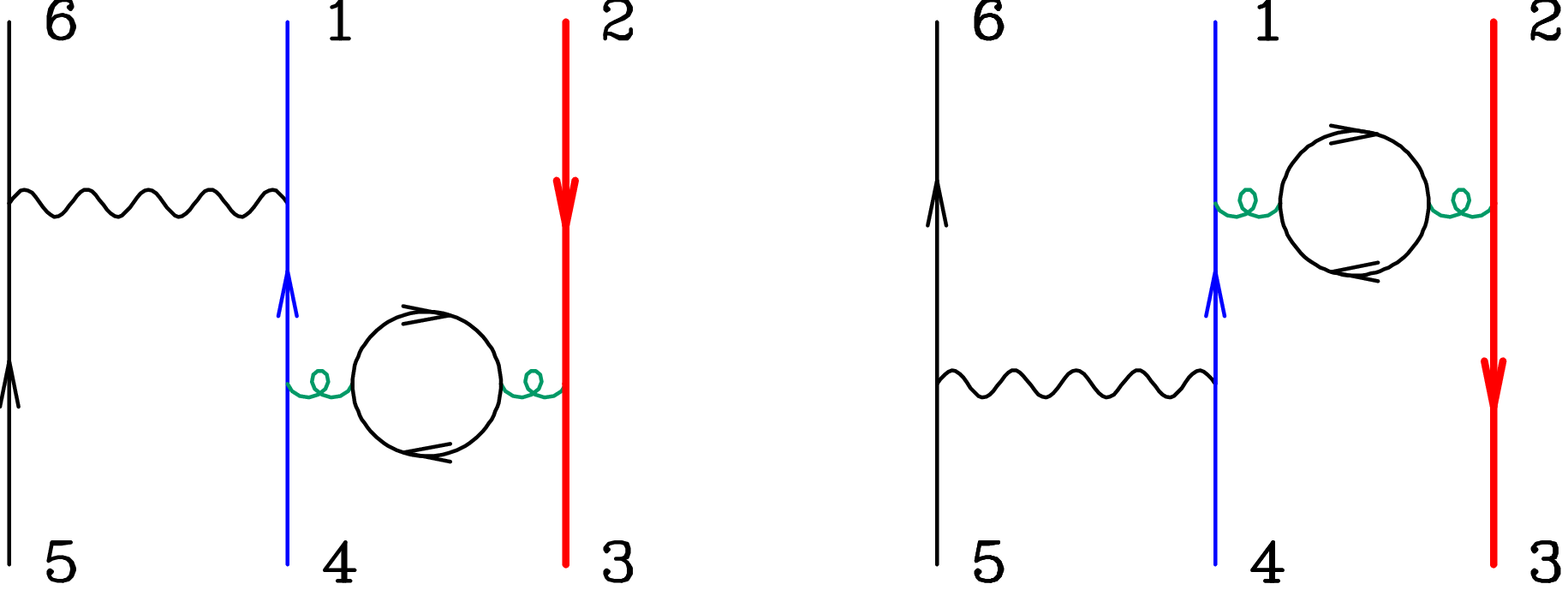}
\end{center}
\caption{Diagrams for the fermion loop primitives $\Alf_6(1,2,3,4)$ and $\Ahf_6(1,2,3,4)$.}
\label{Af}
\end{figure}

Following Ref.~\cite{Bern:1997sc} we can decompose the full one-loop amplitude into gauge invariant
primitive amplitudes,
\begin{eqnarray}
A_{6;1}(1_\q,2_\Qb,3_\Q,4_\qb) &=&\Alc_6(1,2,3,4)-\frac{2}{N^2} (\Acb_6(1,2,3,4)+\Alc_6(1,2,3,4)) \nn \\
   &-&\frac{1}{N^2} \Asl_6(1,2,3,4)-\frac{n_{\lf}}{N}\Alf_6(1,2,3,4)-\frac{n_{\hf}}{N}\Ahf_6(1,2,3,4) \;,
   \nn \\
A_{6;2}(1_\q,2_\Qb,3_\Q,4_\qb) &=& \Acb_6(1,2,3,4)+\frac{1}{N^2} (\Alc_6(1,2,3,4)+\Acb_6(1,2,3,4)) \nn \\
&+&\frac{1}{N^2}\Asl_6(1,2,3,4)+\frac{n_{\lf}}{N}\Alf_6(1,2,3,4)+\frac{n_{\hf}}{N}\Ahf_6(1,2,3,4) \;.
\label{Oneloopcolourdecomp}
\end{eqnarray}
$\Alc_6$ is the primitive amplitude containing the leading colour box and $\Acb_6$ is the primitive amplitude 
containing the crossed box. The parent diagrams are shown in Fig.~\ref{ABLC}. 
These two primitive amplitudes are related by,
\begin{equation}
\Acb_6(1_\q^{h_1},2_\Qb^{h_2},3_\Q^{h_3},4_\qb^{h_4}) = -\Alc_6(1_\q^{h_1},3_\Qb^{h_3},2_\Q^{h_2},4_\qb^{h_4}) \;.
\end{equation}
$\Asl_6$ is the primitive amplitude containing the subleading colour boxes, see Fig.~\ref{Asl},
and the primitive amplitude
for the fermion-loop diagrams (light and heavy), $\Alf_6,\Ahf_6$ is shown in Fig.~\ref{Af}. 
We note that, just as at tree level, parity ensures that we need only compute 
a reduced set of amplitudes. For $\Alc_6$ at one loop it is sufficient to calculate 
three of the four possible spin label combinations since (c.f. Eq.~\ref{pp_from_mm}),
\beq
\Alc_6(1_\q^-,2_\Qb^+, 3_\Q^+,4_\qb^+,5_\ellb^+,6_\ell^-)
 = -{\rm flip} \left[\Alc_6(1_\q^-,2_\Qb^-, 3_\Q^-,4_\qb^+,5_\ellb^+,6_\ell^-) \right] \;.
\label{pp_from_mmloop}
\eeq

\subsection{Structure of the calculation}
In the presence of propagators with vanishing masses on internal lines the
one-loop amplitude contains infrared and collinear divergences. 
In addition, the amplitude contains ultraviolet divergences.  We regulate all of these
divergences using dimensional regularization, continuing the loop
integration to $D=4-2 \e$ dimensions. The divergences then appear as
poles in $\e$.  For the primitive amplitudes the divergence
structure is quite simple and we can separate the amplitude as
follows,
\begin{equation}
\label{VFdecomp}
A_6^{j} = \Bigl[ \Atree_6 V^j + i \, F^j \Bigr]\ ,
\end{equation}
where $V^j$ contains all the divergent pieces.
In this equation $A_6^{j}$ is any of the primitive amplitudes
and the index $j$ runs over $\{\lc,\cb,\sl,\lf,\hf\}$. 
As we shall see later the simplicity of the pole structure gives useful
constraints on the different terms contributing to the amplitude.

\begin{figure}
\begin{center}
\includegraphics[angle=270,scale=0.6]{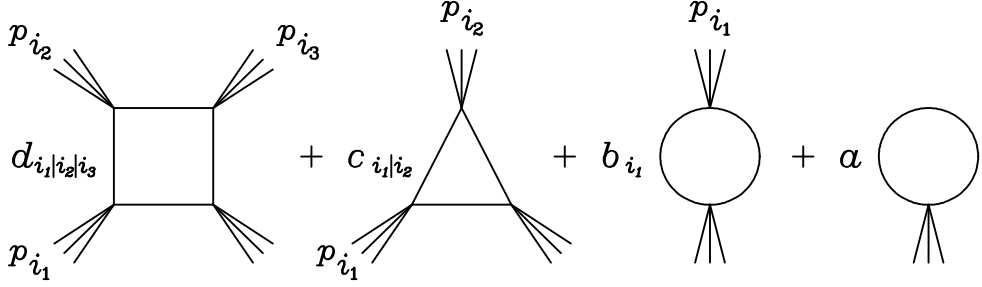} 
\end{center}
\caption{Schematic diagram illustrating the decomposition in scalar integrals.
\label{decomposition}}
\end{figure}
The calculation proceeds by noting that the primitive amplitudes 
can be expressed as a sum of scalar box, triangle, bubble and tadpole
integrals and a rational part,
\beqn \label{scalarintegraldecomp}
&&-i A_6^{j}(1^-,2^{h_2},3^{h_3},4^+,5^+,6^-)= \sum_{i_1 i_2 i_3} d_{i_1|i_2|i_3}(h_3,h_2) \; D_0(p_{i_1},p_{i_2},p_{i_3}) \nn \\
&+&\sum_{i_1 i_2} c_{i_1|i_2}(h_3,h_2) \; C_0(p_{i_1},p_{i_2}) 
+\sum_{i_1} b_{i_1}(h_3,h_2) \; B_0(p_{i_1})+ a(h_3,h_2)\; A_0 \nn \\
&+&R^{j}(1^-,2^{h_2},3^{h_3},4^+,5^+,6^-) \; ,
\eeqn
where the label $j$ runs over the values $\{\lc,\cb,\sl,\lf,\hf\}$.
In this equation, illustrated in Fig.~\ref{decomposition}, the indices $i_1,i_2,i_3$ run over all partitions
of the cyclically ordered momenta. The coefficients $d,c,b,a$ of the scalar integrals in Eq.~(\ref{scalarintegraldecomp})
also depend on the primitive amplitude index $j$. In the following we shall suppress 
this dependence, but the primitive amplitude to which we are referring will be clear from the context. 
The functions $A_0$,~$B_0$,~$C_0$,~$D_0$ are the scalar integrals defined in
Appendix~\ref{intdef}.  The mass labels have been suppressed in the
scalar integrals in Eq.~(\ref{scalarintegraldecomp}) because they are
determined automatically by the propagators connecting the two
external massive quarks.  Values for these scalar integrals are
obtained using the QCDLoop program~\cite{Ellis:2007qk}.  
By using four-dimensional unitarity
methods~\cite{Britto:2004nc,Cachazo:2004kj,Britto:2005ha,Britto:2006sj,Forde:2007mi},
we obtain the coefficients, $a$,~$b$,~$c$,~$d$ in the four-dimensional helicity (FDH)
scheme~\cite{Bern:2002zk}. The rational terms require information from beyond
four dimensions.

\section{The results for primitive amplitude $\Alc_6$}
\subsection{Divergent parts}
For the leading colour primitive amplitude the divergent term which enters in 
Eq.~(\ref{VFdecomp}) is given by~\cite{Catani:2000ef},
\begin{equation}
V^{lc} =-\frac{1}{\e^2} \Bigg[ \Big( \frac{\mu^2}{-2 k_4.p_3}\Big)^\e 
                         + \Big( \frac{\mu^2}{-2 k_1.p_2}\Big)^\e
                         -\Big( \frac{\mu^2}{m^2}\Big)^\e\Bigg]
   +\frac{8}{3} \frac{1}{\e} \;.
\label{Vlc}
\end{equation}
\def\BAB#1.#2.#3.#4{[#1|#2#3|#4]}
\def\ABA#1.#2.#3.#4{\langle#1|#2#3|#4\rangle}
\def\BABAB#1.#2.#3.#4.#5.#6{[#1|#2#3#4#5|#6]}
\def\ABABA#1.#2.#3.#4.#5.#6{\langle#1|#2#3#4#5|#6\rangle}
\def\lo{l_1}
\def\ltt{l_{23}}
\def\lott{l_{123}}

\subsection{Calculation of the box coefficients}
The five scalar box integrals that enter the leading colour amplitude are enumerated
in Table~\ref{boxnot}.
\begin{table}
\begin{tabular}{|l|c||l|c|}
\hline
~~~~Scalar integral              & Coefficient   & ~~~~Scalar integral              & Coefficient \\
\hline
1. $D_0(k_1,p_2,p_3;0,0,m,0)$    & $d_{1|2|3}$   & 4. $D_0(p_2,p_3,k_4;0,m,0,0)$    & $d_{2|3|4}$  \\
2. $D_0(k_1,p_2,p_{34};0,0,m,0)$ & $d_{1|2|34}$  & 5. $D_0(p_{12},p_3,k_4;0,m,0,0)$ & $d_{12|3|4}$ \\
3. $D_0(k_1,p_{23},k_4;0,0,0,0)$ & $d_{1|23|4}$  & & \\
\hline
\end{tabular}
\caption{Scalar box integrals appearing in the leading colour amplitude $A_6^{lc}1,2,3,4)$ and the
notation used in the text to denote their coefficients.
\label{boxnot}}
\end{table}
Of these, we directly calculate the coefficients of boxes 1--3
and obtain the remaining two by symmetry:
\beqn
d_{2|3|4}(h_3, h_2)  &=& -\mbox{flip} \left[ d_{1|2|3}(-h_2, -h_3)\right] \;, \nn \\
d_{12|3|4}(h_3, h_2) &=& -\mbox{flip} \left[ d_{1|2|34}(-h_2, -h_3)\right] \;,
\eeqn
where $h_3$ and $h_2$ represent the spin labels of the quark and antiquark respectively and the
operation $\mbox{flip}$ is defined in Eq.~(\ref{flip1def}).
When using the Rodrigo choice to decompose $p_2$ and $p_3$, box 3
only gives a non-zero contribution for the spin-labels $\{-,-\}$ and $\{+,+\}$.

The coefficients are computed using quadruple cuts with complex
momenta~\cite{Britto:2004nc}. For boxes 1 and 2 the presence of a massive
propagator leads to a more complicated parametrization of the loop momentum
than in the purely massless case. The particular parameterizations that we have
used are spelled out below. 

\subsubsection{Calculation of $d_{1|2|3}$}
\begin{figure}
\begin{center}
\includegraphics[angle=270,scale=0.5]{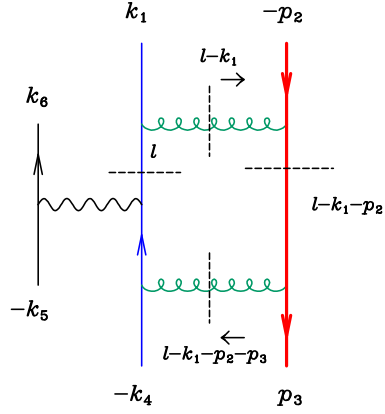}
\end{center}
\caption{Diagram for the calculation of the coefficient of the
scalar integral $D_0(k_1,p_2,p_3;0,0,m,0)$. The massive quark is 
represented by the heavy (red) line and unitarity cuts are 
represented by a dashed line.}
\label{BoxLC321}
\end{figure}

The cuts used to isolate this coefficient are depicted in Fig.~\ref{BoxLC321},
where we remind the reader that we work in the case that $k_1^2=k_4^2=0$ and $p_2^2=p_3^3=m^2$.
The loop momentum is subject to the following constraints,
\beq
l^2 =0,\;\; (l-k_1)^2=0,\;\;(l-k_1-p_2)^2=m^2,\;\;(l-k_1-p_2-p_3)^2=0,\;\;
\label{LC321constraints}
\eeq
Denoting $\lo=l-k_1$ we have $\lo\cdot k_1=0$ so we must have either $\keta{\lo} \sim \keta{1}$ or
$\ketb{\lo} \sim \ketb{1}$, depending on the helicity of the internal gluons,
in such a way that the 3-point $q\bar q g$ vertex is non-vanishing. 

In the case $[\lo| \sim [1|$ we can write 
\beqn
 \lo^{\mu}=l^\mu -k_1^\mu &=& 
\frac{s_{23}}{2} \;  
\frac{\BAB1.\gamma^\mu.\slsh{{\bf 2}}.1}
{\BAB1.(\slsh{{\bf 2}}+\slsh{{\bf 3}}).\slsh{{\bf 2}}.1}
\equiv
\frac{s_{23}}{2} \;  
\frac{\BAB1.\gamma^\mu.\slsh{{\bf 2}}.1}
{\BAB1.\slsh{{\bf 3}}.\slsh{{\bf 2}}.1} \;, \nn \\
 l^{\mu}_{123}=\lo^{\mu}-p_2^\mu-p_3^\mu&=&
-\frac{1}{2} \;  
\frac{\BABAB1.(\slsh{{\bf 2}}+\slsh{{\bf 3}}).\gamma^\mu.(\slsh{{\bf 2}}+\slsh{{\bf 3}}).\slsh{{\bf 2}}.1}
{\BAB1.\slsh{{\bf 3}}.\slsh{{\bf 2}}.1} \;,
\label{LC321square}
\eeqn
where the boldface momenta denote massive vectors.
By inspection, it is clear that this parametrization of the loop
momentum $l^\mu$ satisfies the constraints in Eq.~(\ref{LC321constraints}).
Correspondingly in the case $\langle \lo| \sim \langle 1|$ we can write 
\beqn
\lo^{\mu}=l^\mu - k_1^\mu &=&
\frac{s_{23}}{2} \;  
\frac{\ABA1.\slsh{{\bf 2}}.\gamma^\mu.1}
{\ABA1.\slsh{{\bf 2}}.(\slsh{{\bf 2}}+\slsh{{\bf 3}}).1}
\equiv
\frac{s_{23}}{2} \;  
\frac{\ABA1.\slsh{{\bf 2}}.\gamma^\mu.1}
{\ABA1.\slsh{{\bf 2}}.\slsh{{\bf 3}}.1} \;, \nn \\
 l^{\mu}_{123}=\lo^{\mu}
-p_2^\mu-p_3^\mu&=&
-\frac{1}{2} \;  
\frac{\ABABA1.\slsh{{\bf 2}}.(\slsh{{\bf 2}}+\slsh{{\bf 3}}).\gamma^\mu.(\slsh{{\bf 2}}+\slsh{{\bf 3}}).1}
{\ABA1.\slsh{{\bf 2}}.\slsh{{\bf 3}}.1} \;.
\label{LC321angle}
\eeqn
Note that the denominator in Eq.~(\ref{LC321angle})
simplifies when expressed in terms of massless momenta,
\beq
{\ABA1.\slsh{{\bf 2}}.\slsh{{\bf 3}}.1} = \beta \spa1.2 \spb2.3 \spa3.1 \;,
\eeq

As an example, we will calculate the coefficient of this box integral for the spin label
choice $\{-,+ \}$. The box coefficient is given by the formula,
\beqn
d_{1|2|3}(-,+) &\equiv& \frac{1}{2} \sum_{h_a, h_b} d_{1|2|3}^{\,[h_a,h_b]}(-,+) \nn \\
&=& \frac{1}{2}
\sum_{h_a, h_b} A_3(1_\q^-, -{l_\qb^+}, \lo^{h_a})
            \times A_5({l_\q^-}, 4_\qb^+, -\lott^{h_b}, 5_\lb^+,6_\l^-) \nn \\
  && \qquad \times A_4(-\lo^{-h_a}, 2_\Qb^+, 3_\Q^-, \lott^{-h_b})
	    \times (-i) \left[ (l-k_1-p_2)^2 - m^2 \right] \;,
\eeqn
with the loop momenta fixed according to the constraints above, i.e.
$l^\mu$ is given by either Eq.~(\ref{LC321square})
or~(\ref{LC321angle}) above.  In this formula we have for brevity
suppressed the gluon particle labels. Note that we have not explicitly
cut the heavy quark propagator but, equivalently, have simply written
the full $g\Qb \Q g$ amplitude and multiplied by the cut
propagator. Since the $g\Qb \Q g$ amplitudes with the same helicity
gluons vanish (c.f. Eq.~(\ref{gQbQgresults})), the coefficient
receives only two contributions.

Let us first inspect the assignment $h_a = -h_b = +1$. We note that it is simplest to
manipulate the $g\Qb \Q g$ amplitude, (Eq.~(\ref{gQbQgresults}) first line), 
using the Schouten identity, Eq.~(\ref{Schouten}), 
into an alternative form in order to take full advantage of the vanishing of the cut propagator,
\beqn
-i A_4(-\lo^-,2_\Qb^+,3_\Q^-,\lott^+) \left[ (l_1-p_2)^2 - m^2 \right]
 &=& \frac{\spb2.\lott^2 \spab \lo.\slsh{\bf 2}.\lott} 
{\spb \lo.\lott \spb2.3 } \nn \\
 &=& -\frac{\spb2.\lott^2
  \left( \spb \lo.\lott \spab \lo.\slsh{\bf 2}.3 - \spab \lo.\slsh{\bf 2}.\lo \spb3.\lott \right)}
{\spb \lo.3 \spb \lo.\lott \spb2.3 } \nn \\
 &=& -\frac{\spb2.\lott^2}{\spb \lo.3 \spb2.3}
  \left( \spab \lo.\slsh{\bf 2}.3
       - \frac{\spb3.\lott \spab \lo.\slsh{\bf 2}.\lo}{\spb \lo.\lott} \right) \;.
\eeqn
On shell, $ (l_1-p_2)^2 - m^2 \equiv \spab \lo.\slsh{\bf 2}.\lo = 0$ 
and we can simply discard the second term in this equation. Using results for the
other amplitudes presented in Appendix~\ref{treeapp} we find that,
\beq
d_{1|2|3}^{\,[+,-]}(-,+) = \frac{\spb {l}.\lo^2}{\spb l.1} \times
\frac{\spb4.5^2}{\spb4.\lott \spb l.\lott \spb6.5} \times
\frac{\beta_+ \spb2.\lott^2 \spa2.\lo}{\spb \lo.3} \;,
\eeq
where we have simplified the $g\Qb \Q g$ amplitude further using the decomposition
of $p_2$ in terms of $k_2$ and $k_3$. Since the propagator $l$ does not
have a simple decomposition in terms of external momenta, it is simplest to
eliminate it by multiplying this expression in numerator and denominator by 
a factor $\spa4.l \spa\lott.l$. After some simplification this yields,
\beq
d_{1|2|3}^{\,[+,-]}(-,+) = 
\frac{\beta_+ \spb4.5^2 \spa4.1}{s_{123} \spb6.5} \,
\frac{\spa\lott.1 \spb1.\lo \spb2.\lott^2 \spa2.\lo}{\spa 4.\lo \spb4.\lott \spb \lo.3} \;,
\eeq
This is now in a form where we can use the loop momenta definitions given in Eq.~(\ref{LC321angle}) with
spinors identified by, for example, using the identity, $l^\mu = \frac{1}{2} \spab l.{\gamma^\mu}.l$. We can
thus make the replacements:
\beq
\setlength\arraycolsep{0.2em}
\begin{array}{lll}
\spa\lott.1 &\longrightarrow 
-\frac{\displaystyle \ABA1.\slsh{{\bf 2}}.(\slsh{{\bf 2}}+\slsh{{\bf 3}}).1}
{\displaystyle \ABA1.\slsh{{\bf 2}}.\slsh{{\bf 3}}.1} = -1 \;, &
\spb1.\lo \longrightarrow 
-s_{23}\frac{\displaystyle \spab1.\slsh{{\bf 2}}.1}
{\displaystyle \ABA1.\slsh{{\bf 2}}.\slsh{{\bf 3}}.1} \;, \\
\spb4.\lott &\longrightarrow
 \spba4.(\slsh{{\bf 2}}+\slsh{{\bf 3}}).1 = \spab1.2+3.4 \;, & 
\spb2.\lott \longrightarrow
 \spba2.(\slsh{{\bf 2}}+\slsh{{\bf 3}}).1 = \spb2.3 \spa3.1 \;, \\
\spa2.\lo &\longrightarrow \spa2.1 \;, &
\spa4.\lo \longrightarrow \spa4.1 \;, \\
\spb \lo.3 &\longrightarrow 
s_{23}\frac{\displaystyle \spab1.\slsh{{\bf 2}}.3}
{\displaystyle \ABA1.\slsh{{\bf 2}}.\slsh{{\bf 3}}.1} = s_{23} \frac{\displaystyle \beta_+ \spa1.2 \spb2.3}
{\displaystyle \ABA1.\slsh{{\bf 2}}.\slsh{{\bf 3}}.1} \;.&
\end{array}
\eeq
This results in the final expression,
\beq
d_{1|2|3}^{\,[+,-]}(-,+) = \frac{\spa1.3^2 \spb2.3 \spb4.5^2 \spab1.\slsh{{\bf 2}}.1}{s_{123} \spab1.2+3.4 \spb5.6} \;.
\eeq
Adding in the contribution from the other helicity configuration
($h_a=-h_b = -1$), computed in a similar fashion, yields the full result
for this coefficient,
\beqn
d_{1|2|3}(-,+) &=&
\frac{s_{23} \spab 1.\slsh{{\bf 2}}.1}{2s_{123}}
 \left[ \frac{\spb 2.3}{\spa 5.6 \spab 4.2+3.1}
 \left( \frac{\spab 6.\slsh{{\bf 2}}.1}{\beta \spb 1.3}	 
      + \frac{\spa 1.6 \spb 1.2}{\spb 2.3}	  
 \right)^2
  - \frac{\spa 1.3^2 \spb 4.5^2}{\spa 2.3 \spb 5.6 \spab 1.2+3.4 }	
 \right] \;. \nn \\
\eeqn
This is in agreement with the coefficient presented in Ref.~\cite{Bern:1997sc} in the limit that the heavy quark mass is taken to zero.

\subsubsection{Calculation of $d_{1|2|34}$}
\begin{figure}
\begin{center}
\includegraphics[angle=270,scale=0.5]{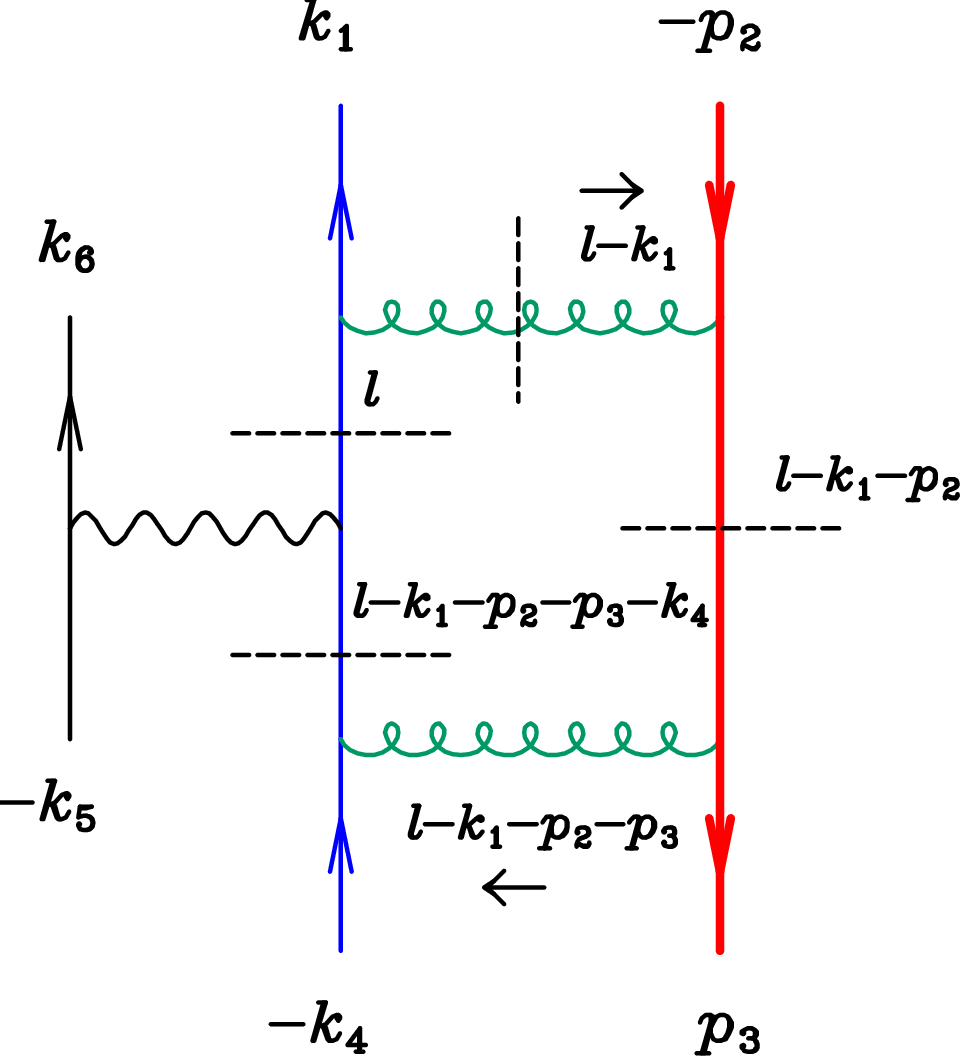}
\end{center}
\caption{Diagram for the calculation of the coefficient of the
scalar integral $D_0(k_1,p_2,p_{34};0,0,m,0)$. The massive quark is 
represented by the heavy (red) line and unitarity cuts are 
represented by a dashed line.}
\label{BoxLC3421}
\end{figure}

This coefficient is obtained from the cuts shown in Fig.~\ref{BoxLC3421},
where the loop momentum is subject to the conditions,
\beq
l^2=0,\;\;
(l-k_1)^2=0,\;\;
(l-k_1-p_2)^2=m^2,\;\;
l_{1234}^2 = (l-k_1-p_2-p_3-k_4)^2=0 \; .
\label{LC3421constraints}
\eeq
An explicit solution for $l_1=l-k_1$ is given by,
\beq
\lo^\mu = \frac{s_{234}}{2} \frac{\ABA1.\slsh{{\bf 2}}.\gamma^\mu.1 }
   {\ABA1.\slsh{{\bf 2}}.(\slsh{{\bf 3}}+\slsh{4}).1} \;,
\label{l1x2x34}
\eeq
which clearly satisfies the constraints in Eq.~(\ref{LC3421constraints}).
From this we can also derive a useful form for the momentum $l_{1234}$,
\beq
l_{1234}^\mu =\lo-p_2-p_3-k_4 =  -\frac{1}{2} \frac{\ABABA 1.
\slsh{{\bf 2}}.{(\slsh{\bf 2}+\slsh{\bf 3}+\slsh{4})}. 
\gamma^\mu.
{(\slsh{\bf 2}+\slsh{\bf 3}+\slsh{4})}.1 }
{\ABA1.\slsh{\bf 2}.{(\slsh{\bf 3}+\slsh{4})}.1} \;.
\eeq
The other solution, corresponding to $\brab{l_1} \sim \brab{1}$ can be obtained from
Eq.~(\ref{l1x2x34}) by complex conjugation. We note that both of these parameterizations
can be obtained from Eqs.~(\ref{LC321square}) and~(\ref{LC321angle}) by simply making the
replacements, $\slsh{\bf 3} \to (\slsh{\bf 3}+\slsh{4})$ and $s_{23} \to s_{234}$.
From these forms it is straightforward to compute the coefficients of this
box.

\subsubsection{Calculation of $d_{1|23|4}$}
\begin{figure}
\begin{center}
\includegraphics[angle=270,scale=0.5]{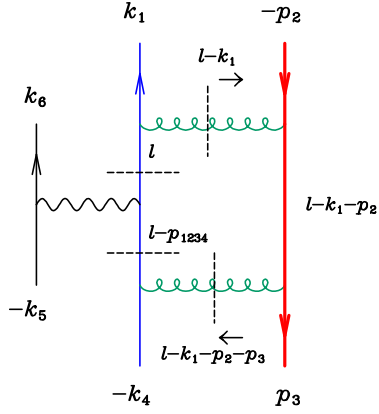}
\end{center}
\caption{Diagram for the calculation of the coefficient of the
scalar integral $D_0(k_1,p_{23},k_4;0,0,0,0)$. The massive quark is 
represented by the heavy (red) line and unitarity cuts are 
represented by a dashed line.}
\label{BoxLC1x23x4}
\end{figure}

The cuts used to isolate this coefficient are depicted in Fig.~\ref{BoxLC1x23x4}.
This box with massless internal lines and two opposite massive external lines
is often referred to as the easy box~\cite{Bern:1993kr}. 
The heavy quark does not participate in the loop, and 
the loop momentum is subject to the following constraints,
\beq
l^2 =0,\;\; (l-k_1)^2=0,\;\;(l-k_1-p_2-p_3)^2=0,\;\;(l-k_1-p_2-p_3-k_4)^2=0,\;\;
\label{LC1x23x4constraints}
\eeq
The explicit solutions for $\lo=l-k_1$ and $l_{123}=l-k_1-p_2-p_3$ are given either by,
\beqn \label{LC1x23x4angle}
\lo^\mu &=& ~~\frac{1}{2} \frac{\ABA4.{{\bf (\slsh{2}+\slsh{3})}}.\gamma^\mu.1 }
   {\spa4.1} \;, \nn \\
l_{123}^\mu &=& -\frac{1}{2} \frac{\ABA4.\gamma^\mu.{{\bf (\slsh{2}+\slsh{3})}}.1 }
   {\spa4.1} \;,
\eeqn
or by,
\beqn \label{LC1x23x4square}
\lo^\mu &=& ~~\frac{1}{2} \frac{\BAB4.{{\bf (\slsh{2}+\slsh{3})}}.\gamma^\mu.1 }
   {\spb4.1} \;, \nn \\
l_{123}^\mu &=& -\frac{1}{2} \frac{\BAB4.\gamma^\mu.{{\bf (\slsh{2}+\slsh{3})}}.1 }
   {\spb4.1} \;.
\eeqn
Both of these clearly satisfy the constraints in Eq.~(\ref{LC1x23x4constraints}).

As an example, we will calculate the coefficient of this box integral for the spin label
choice $\{-,+\}$. For this spin label choice it turns out that the box 
coefficient vanishes. 
The argument is quite general and may be useful in explaining the absence 
of the easy box in other contexts, e.g.\ ref.~\cite{Badger:2009vh}. 
The result for the box coefficient is calculated from,
\beqn
d_{1|23|4}(-,+) =
\sum_{h_a, h_b} && A_3(1_\q^-, -{l^+}_\qb, \lo^{h_a})
            \times A_4({l^-}_\q, -l_{1234\;\qb}^+, 5_\lb^+,6_\l^-) \nn \\
         && \times A_3(l_{1234\; \q}^-, 4_\qb^+, -\l_{123}^{h_b}) \times A_4(-\lo^{-h_a}, 2_\Qb^+, 3_\Q^-, \lott^{-h_b})\;,
\eeqn
with the loop momenta fixed according to the constraints above, i.e.
$l^\mu$ is given by either Eq.~(\ref{LC1x23x4angle}) or~(\ref{LC1x23x4square}) above.
From Eq.~(\ref{gQbQgresults}) we see that $A_4(-\lo^{-h_a}, 2_\Qb^+, 3_\Q^-, \lott^{-h_b})$ vanishes 
for $h_a=h_b$. Therefore we need only consider the case $h_a=-h_b$. For definiteness let 
us consider the case $h_a=1,h_b=-1$.
For this case we find that,
\beqn
\label{explicit3vertices}
A_3(1_\q^-, -{l^+}_\qb, \lo^+) = \frac{{\spb l.\lo}^2}{\spb1.l \spb \lo.1}
 \quad &\implies& |\lo \ra \sim |1\ra \;, \nn \\
A_3(l_{1234\; \q}^-, 4_\qb^+, -\l_{123}^-) = \frac{{\spa l_{1234}.\lott}^2}{\spa4.\lott \spa 4.{l_{1234}}}
 \quad &\implies& |\l_{123}] \sim |4] \;.
\eeqn
However we see that the conditions $|\lo \ra \sim |1\ra, |\l_{123}] \sim |4]$ in Eq.~(\ref{explicit3vertices}) 
are not compatible with the  kinematic constraints, Eqs.~(\ref{LC1x23x4angle},\ref{LC1x23x4square}) which require either,
\beqn
&|\lo \ra \sim |1\ra~\mbox{and}~&|\l_{123}\ra \sim |4\ra \;, \nn \\
{\rm or} &|\lo ] \sim |1]~\mbox{and}~&|\l_{123}] \sim |4] \;,
\eeqn
respectively. The argument follows identically for the contribution  $h_a=-1,h_b=+1$.
Therefore for this external spin label choice $\{-,+\}$ the box coefficient is zero. 
For the spin label choice $\{-,-\}$ there is a non-zero easy box coefficient because
$A_4(-\lo^{-h_a}, 2_\Qb^+, 3_\Q^-, \lott^{-h_b})$ no longer vanishes for $h_a=h_b$.
\def\bm{\beta_-}
\def\bp{\beta_+}
\def\be{\beta}

\subsection{Calculation of triangle coefficients}
The leading colour amplitude receives contributions from the ten triangle scalar integrals
listed in Table~\ref{trinot}.
\begin{table}
\begin{tabular}{|l|c||l|c|}
\hline
~~~~Scalar integral           & Coefficient & ~~~~Scalar integral          & Coefficient \\
\hline
1. $C_0(p_{23},k_4;0,0,0)$    & $c_{23|4}$  & 6.  $C_0(k_1,p_{23},0,0,0)$  & $c_{1|23}$ \\
2. $C_0(p_{12},p_3;0,m,0)$    & $c_{12|3}$  & 7.  $C_0(p_3,k_4,m,0,0)$     & $c_{3|4}$ \\
3. $C_0(p_{12},p_{34};0,m,0)$ & $c_{12|34}$ & 8.  $C_0(k_1,p_2,0,0,m)$     & $c_{1|2}$ \\
4. $C_0(p_2,p_3;0,m,0)$       & $c_{2|3}$   & 9.  $C_0(k_1,p_{234},0,0,0)$ & $c_{1|234}$ \\
5. $C_0(p_2,p_{34};0,m,0)$    & $c_{2|34}$  & 10. $C_0(p_{123},k_4,0,0,0)$ & $c_{123|4}$ \\
\hline
\end{tabular}
\caption{Scalar triangle integrals appearing in the leading colour amplitude $\Alc_6(1,2,3,4)$ and the
notation used in the text to denote their coefficients.
\label{trinot}}
\end{table}
The coefficients of Triangles 1--4 are calculated directly 
as detailed in the subsection below. 
Triangle 5 can be simply obtained by symmetry,
\beq
c_{2|34}(h_3, h_2) = -\mbox{flip} \left[ c_{12|3}(-h_2, -h_3)\right] \;. 
\eeq

The coefficients of the five remaining triangles are then uniquely determined by
the known divergence structure~\cite{Catani:2000ef} of the amplitude, Eq.~(\ref{Vlc}). 
The terms in Eq.~(\ref{Vlc}) proportional to $\log(-2k_1.p_2)/\e$ and
$\log(-2k_4.p_3)/\e$ fix two of the triangle coefficients,
\beqn
c_{1|2} &=& - \left( \spab1.{\slsh{\bf 2}}.1 \left(-i A^{{\rm tree}}_6\right) + \frac{d_{1|2|3}}{s_{23}}
 + \frac{d_{1|2|34}}{s_{234}} \right) \;, \nn \\
c_{3|4} &=& - \left( \spab4.{\slsh{\bf 3}}.4 \left(-i A^{{\rm tree}}_6\right) + \frac{d_{2|3|4}}{s_{23}}
 + \frac{d_{12|3|4}}{s_{123}} \right) \;,
\eeqn
while the absence of single poles of the form $\log(s)/\e$ for
$s \in \{s_{23},s_{1234},s_{123},s_{234}\}$ requires the following relations,
\beqn
c_{1|23} &=& \spab1.{2+3}.1\left( \frac{d_{1|2|3}}{s_{23} \spab1.{\slsh{\bf 2}}.1} + \frac{d_{2|3|4}}{s_{23} \spab4.{\slsh{\bf 3}}.4}
 - \frac{c_{23|4}}{\spab4.{2+3}.4} - \frac{2 d_{1|23|4}}{\spab1.{2+3}.4 \spab4.{2+3}.1} \right) \;, \nn \\
c_{1|234} &=& \spab1.{2+3+4}.1 \left( \frac{d_{1|2|3}}{s_{23} \spab1.{\slsh{\bf 2}}.1}
 + \frac{d_{1|2|34}}{s_{234} \spab1.{\slsh{\bf 2}}.1} - \frac{c_{1|23}}{\spab1.{2+3}.1} \right) \;, \\
c_{123|4} &=& \spab4.{1+2+3}.4 \left( \frac{d_{12|3|4}}{s_{123} \spab4.{\slsh{\bf 3}}.4}
 + \frac{d_{1|2|34}}{s_{234} \spab1.{\slsh{\bf 2}}.1} 
 - \frac{c_{1|234}}{\spab1.{2+3+4}.1} + \frac{2 d_{1|23|4}}{\spab1.{2+3}.4 \spab4.{2+3}.1} \right) \;. \nn
\eeqn
These expressions are written in terms of already-calculated box and triangle coefficients,
the leading order amplitude, and the other triangle coefficients discussed below.

\subsubsection{Forde method for triangle coefficients}

\begin{figure}
\begin{center}
\includegraphics[angle=270,scale=0.5]{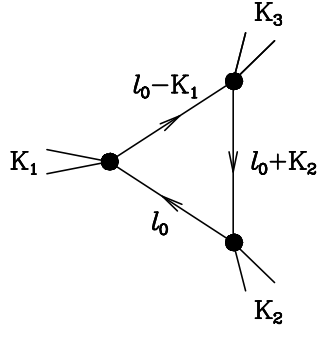}
\end{center}
\caption{Triangle diagram showing the momentum parameterization. All momenta are outgoing, $K_1+K_2+K_3=0$.}
\label{triangle}
\end{figure}

We will calculate the coefficients of the triangle integrals
using the method of Forde~\cite{Forde:2007mi}. 
Triangles 2, 3 and 4 will require 
a slight extension of the formalism to include
one of the internal propagators with a mass.
We first review the case 
of three massless internal momenta, shown in 
Fig.~\ref{triangle} in order to introduce our notation which differs from that of Forde.
Defining $l_1$ and $l_2$ as follows,  
\begin{eqnarray}
l_1^\mu = l_0^\mu-K_1^\mu \;, 
&&
l_2^\mu = l_0^\mu+K_2^\mu \;,
\end{eqnarray}
the cut loop momenta $(l_i^2=0), i=0,1,2$ may be written in the following general form,
\begin{eqnarray} \label{lexpansion}
l^{\mu}_i=x_{i} K_1^{\flat,\mu}+y_{i} K_2^{\flat,\mu}+\frac{t}{2}\spab{K^{\flat}_1}.{\gamma^{\mu}}.{K^{\flat}_2}
+\frac{x_{i}y_{i}}{2t}\spab{K^{\flat}_2}.{\gamma^{\mu}}.{K^{\flat}_1} \;.
\end{eqnarray}
All momenta can be expanded in terms of massless momenta, $K_1^\flat$ and $K_2^\flat$, 
\begin{eqnarray} \label{flattery}
K_1 &=& K_{1}^{\flat}+\frac{S_1}{\gamma}  K_{2}^{\flat} \; ,\nn \\
K_2 &=& K_{2}^{\flat}+\frac{S_2}{\gamma}  K_{1}^{\flat} \; ,\nn \\
K_3 &=& -(1+\frac{S_2}{\gamma}) K_{1}^{\flat}-(1+\frac{S_1}{\gamma}) K_{2}^{\flat} \; ,
\end{eqnarray}
where $S_i=K_i^2$ and $\gamma = \spab{K_{1}^{\flat}}.{K_{2}^{\flat}}.{K_{1}^{\flat}} = 2 {K_{2}^{\flat}}\cdot{K_{1}^{\flat}} $.
The inverse relations are, 
\begin{eqnarray} \label{invflattery}
K_1^{\flat,\mu}=\frac{K^{\mu}_1-(S_1/\gamma)K_2^{\mu}}{1-(S_1S_2/\gamma^2)},
\;\;\;\;\;K_2^{\flat,\mu}=\frac{K_2^{\mu}-(S_2/\gamma)K^{\mu}_1}{1-(S_1S_2/
\gamma^2)}\; .
\end{eqnarray}
The dot product of $K_1$ and $K_2$ in Eq.~(\ref{flattery})
produces a quadratic equation for $\gamma$,
the solutions of which express $\gamma$  in terms of the external momenta,
\begin{eqnarray}
\gamma_{\pm}=(K_1\cdot K_2)\pm\sqrt{\Delta},\;\;\;\;\;\;\Delta=(K_1\cdot K_2)^2-S_1 S_2 \; .
\end{eqnarray}
From Eq.~(\ref{lexpansion}) the massless vectors $l_i$ can be expressed 
as a linear combination 
of the spinor solutions for the vectors $K_{1}^{\flat}$ and $K_{2}^{\flat}$,
\begin{eqnarray} \label{lexpandedinspinors}
\langle l_i|
=t\langle K_{1}^{\flat}|+y_{i}\langle K_{2}^{\flat}| \;,
&&
[ l_i|=\frac{x_{i}}{t} [K_{1}^{\flat}|+[ K_{2}^{\flat}| \;.
\end{eqnarray}
The on-shell conditions $l_i^2=0$ for $i=0,1,2$ allow us to derive the coefficients, $x_{i}$ and $y_{i}$,
\begin{eqnarray}
y_0=\frac{S_1\left(\gamma+S_2\right)}{\left(\gamma^2-S_1S_2\right)},\;\;&&\;\;x_0=-\frac{S_2\left(\gamma+S_1\right)}{\left(\gamma^2-S_1S_2\right)},
\nonumber\\
y_1=y_0-\frac{S_1}{\gamma}=\frac{S_1S_2\left(\gamma+S_1\right)}{\gamma(\gamma^2-S_1S_2)},\;\;&&\;\;
x_1=x_0-1=-\frac{\gamma(\gamma+S_2)}{\gamma^2-S_1S_2},
\nonumber\\
y_2=y_0+1=\frac{\gamma(\gamma+S_1)}{\gamma^2-S_1S_2},\;\;&&\;\;
x_2=x_0+\frac{S_2}{\gamma}=-\frac{S_1S_2\left(\gamma+S_2\right)}{\gamma(\gamma^2-S_1S_2)} \; .
\end{eqnarray}

The spinor products can be expressed as follows,
\begin{eqnarray}
\left[ll_1\right]&=&\frac{x_1-x_0}{t}[K^{\flat}_2
  K^{\flat}_1]=-\frac{1}{t} [K^{\flat}_2K^{\flat}_1],
\nonumber\\
\spa{l}.{l_1}&=&t(y_1-y_0)\spa{K^{\flat}_1}.{K^{\flat}_2}=
 -\frac{tS_1}{\gamma}\spa{K^{\flat}_1}.{K^{\flat}_2},
\nonumber\\
\left[ll_2\right]&=&\frac{x_2-x_0}{t}[K^{\flat}_2
  K^{\flat}_1]=\frac{S_2}{\gamma t}[K^{\flat}_2
  K^{\flat}_1],
\nonumber\\
\spa{l}.{l_2}&=&t(y_2-y_0)\spa{K^{\flat}_1}.{K^{\flat}_2}=
  t\spa{K^{\flat}_1}.{K^{\flat}_2},
\nonumber\\
\left[l_1l_2\right]&=&\frac{x_2-x_1}{t}[K^{\flat}_2K^{\flat}_1]=\frac{1}{t}\left(1+\frac{S_2}{\gamma}\right)[K^{\flat}_2K^{\flat}_1],
\nonumber\\
\spa{l_1}.{l_2}&=&t(y_2-y_1)\spa{K^{\flat}_1}.{K^{\flat}_2}=t\left(1+\frac{S_1}{\gamma}\right)\spa{K^{\flat}_1}.{K^{\flat}_2}.
\end{eqnarray}

Turning now to the case where the propagator with momentum $l_2$ has a mass,
we have that $l_0^2=l_1^2=0$ and $l_2^2=m^2$. 
Since $l_2$ is no longer massless it does not have an expansion of the form Eq.~(\ref{lexpandedinspinors}).
The results for the coefficients 
in the expansion defined in Eq.~(\ref{lexpandedinspinors}) for  $l_0$ and $l_1$ are,
\begin{eqnarray}
y_0=\frac{S_1\left(\gamma+S_2-m^2\right)}{\left(\gamma^2-S_1S_2\right)},\;\;&&\;\;
x_0=\frac{\gamma m^2-S_2\left(\gamma+S_1\right)}{\left(\gamma^2-S_1S_2\right)},
\nonumber\\
y_1=y_0-\frac{S_1}{\gamma}=\frac{S_1\left(S_2\left(1+(S_1/\gamma)\right)-m^2\right)}{\gamma^2-S_1S_2},\;\;&&\;\;
x_1=x_0-1=\frac{\gamma m^2-\gamma(S_2+\gamma)}{\gamma^2-S_1S_2} \; .
\end{eqnarray}
\begin{figure}
\begin{center}
\includegraphics[angle=270,scale=0.5]{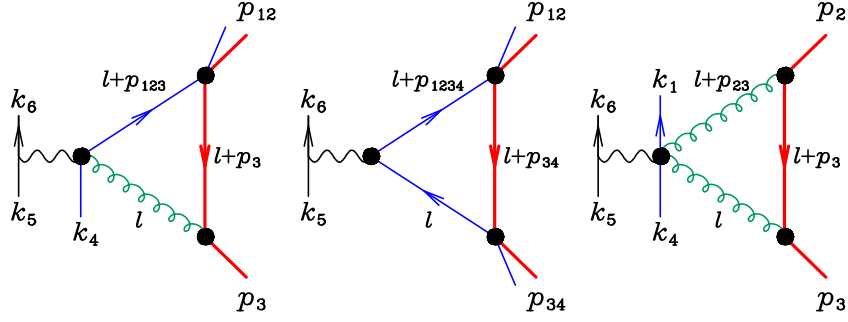}
\end{center}
\caption{Diagrams for the calculation of the coefficients of the scalar integrals
$C_0(p_{12},p_3;0,m,0)$, $C_0(p_{12},p_{34};0,m,0)$ and $C_0(p_{2},p_3;0,m,0)$. The massive quark is 
represented by the heavy (red) line. The momenta on the external lines are all outgoing.}
\label{triangle1}
\end{figure}

With these results in hand we can then compute the coefficients for triangles 2--4 with the loop
momentum assignments as shown in Figure~\ref{triangle1} by making the replacements:
\begin{itemize}
\item Triangle 2: $K_1=-p_{123}$, $K_2=p_3$.
\item Triangle 3: $K_1=-p_{1234}$, $K_2=p_{34}$.
\item Triangle 4: $K_1=-p_{23}$, $K_2=p_{3}$.
\end{itemize}
For the case of triangles 2 and 4, we see that the 
coefficients are particularly simple
since $S_2 = m^2$.

We shall sketch the calculation of the coefficient of triangle 4 because it is the simplest.
We have $S_1=p_{23}^2=s_{23}$ and $S_2=p_3^2=m^2$ and hence see that,
\beq
\gamma_{\pm} = K_1 \cdot K_2 \pm \sqrt{(K_1 \cdot K_2)^2 - S_1 S_2} = 
 -\beta_{\mp} s_{23}\; .
\eeq
The lightlike momenta, Eq.~(\ref{invflattery}), for the two solutions for $\gamma$ reduce to,
\beqn
&& K^{\flat}_1 = -k_2 \;, \quad K^{\flat}_2 = \bp k_3 \;, \quad 
   {\rm for}~\gamma=\gamma_{-}= -\beta_{+} s_{23}, \; y_0 = -\frac{1}{\beta},\;x_0 =-\frac{\beta_-}{\beta} \;, \nn \\
&& K^{\flat}_1 = -k_3 \;, \quad K^{\flat}_2 = \bm k_2 \;, \quad 
   {\rm for}~\gamma=\gamma_{+}= -\beta_{-} s_{23}, \; y_0 =+\frac{1}{\beta}, \;x_0 =+\frac{\beta_+}{\beta} \;.
\eeqn
Using these assignments we find that the formulae for the coefficients of this triangle are
relatively compact. For example, for the $\{-,+\}$ spin labels the result is,
\beqn
&& c_{2|3}(-,+) = \frac{1}{2s_{56}} \Bigg[ \Biggl\{
\frac{\spb4.5}{s_{123}} \Bigl( 
-\frac{\bm\spa1.3^2\spab6.(1+2).3}{\spa1.2}
-\frac{\bp^2\spa1.2\spb1.2^2\spab6.(1+2).3}{\be \spb1.3^2} \nn \\ &&
+\frac{\bp\bm\spa1.2\spb1.2\spab6.(1+3).2}{\be \spb1.3} 
-\frac{\bp\bm\spa1.3\spb1.2\spab6.(1+2).3}{\be \spb1.3} \nn \\ &&
+\frac{(\be^2-\bm\bp\be+4\bm\bp^2)}{\be} \spa1.3\spab6.(1+3).2 \Bigr) 
+\frac{\bm\spa1.3\spa1.6\spb4.5}{\spa1.2} \nn \\ &&
-\frac{\bp^2\spa2.3\spab4.(2+3).5\spab6.(1+2).3}{\be\spa2.4^2\spb1.3}
+\frac{\bp\spab3.(2+4).5\spab6.(1+2).3}{\spa2.4\spb1.3}\Biggr\}  -\Biggl\{ \mbox{flip}
\Biggr\}\Bigg]\; .
\eeqn

\subsection{Calculation of bubble and tadpole coefficients}
\label{bubblesec}
The bubble integrals present in the leading colour amplitude are shown in Table~\ref{bubnot}.
\begin{table}
\begin{tabular}{|l|c||l|c|}
\hline
~~~~Scalar integral    & Coefficient & ~~~~Scalar integral    & Coefficient \\
\hline
1. $B_0(p_{1234};0,0)$ & $b_{1234}$  & 4. $B_0(p_{12};0,m)$   & $b_{12}$  \\
2. $B_0(p_{23};0,0)$   & $b_{23}$    & 5. $B_0(p_{34};0,m)$   & $b_{34}$  \\
3. $B_0(p_{123};0,0)$  & $b_{123}$   & 6. $B_0(p_{234};0,0)$  & $b_{234}$ \\
\hline
\hline
~~~~$B_0(p_2;0,m)$     & $b_{2}$     & ~~~~$A_0(m)$           & $a$ \\
\hline
\end{tabular}
\caption{Scalar bubble and tadpole integrals appearing in the leading colour amplitude
$\Alc_6(1,2,3,4)$ and the notation used in the text to denote their coefficients.
\label{bubnot}}
\end{table}
The coefficients of bubbles 1--4 are computed by using the method of spinor
integration~\cite{Cachazo:2004kj,Britto:2005ha,Britto:2006sj}. This is straightforward
to apply for bubbles 1--3 but requires a small modification for bubble 4 due to the effect
of the massive propagator. The modified method is described in the subsection below. 
Coefficients of the bubble integrals 5 and 6 are then obtained by symmetry,
\beqn
b_{34}(h_3, h_2)  &=& -\mbox{flip} \left[ b_{12}(-h_2, -h_3)\right] \;, \nn \\
b_{234}(h_3, h_2) &=& -\mbox{flip} \left[ b_{123}(-h_2, -h_3)\right] \;,
\eeqn

All occurrences of $B_0(p_3;0,m)$ have been replaced by $B_0(p_2;0,m)$, so we only 
need to determine the coefficient $b_2$.
A linear combination of the coefficient $b_2$ and $a$ is determined from the known form
of the single pole in $\e$, (i.e.\ the single pole 
with no associated logarithm in the expansion of Eq.~(\ref{Vlc})),
\beq \label{bsum}
b_{2}+m^2 a = \frac{8}{3} (-i) A^{{\rm tree}}_6 - b_{123} - b_{234} - b_{23} - b_{1234} - b_{12} - b_{34} \;. 
\eeq
Eq.~(\ref{bsum}) is sufficient to fix the poles and the logarithms, but since,
\beqn
A_0(m)/m^2 &=&  \Big( \frac{\mu^2}{m^2}\Big)^\e \Big( \frac{1}{\e} +1\Big) \;, \nn \\ 
B_0(p_2;0,m)&=& \Big( \frac{\mu^2}{m^2}\Big)^\e \Big( \frac{1}{\e} +2\Big) \;,
\eeqn
it leaves the constant term undetermined.
In order to separately fix the coefficients $a$ and $b_2$
we perform a direct Feynman diagram computation of the tadpole coefficient $a$ to find,
\beqn
a(-,+) &=& 0 \; ,  \nn \\
a(+,-) &=& 0 \; ,  \nn \\
m^2 a(-,-) &=& \frac{1}{2} (-i) A^{{\rm tree}}_6(1_\q^-,2_\Qb^-,3_\Q^-,4_\qb^+,5_\ellb^+,6_\ell^-) \; ,\nn \\
m^2 a(+,+) &=& \frac{1}{2} (-i) A^{{\rm tree}}_6(1_\q^-,2_\Qb^+,3_\Q^+,4_\qb^+,5_\ellb^+,6_\ell^-) \; .
\eeqn
These results for $a$ are valid in any covariant gauge.

\subsubsection{Bubble integral with one massive propagator}
\begin{figure}
\begin{center}
\includegraphics[angle=270,scale=0.5]{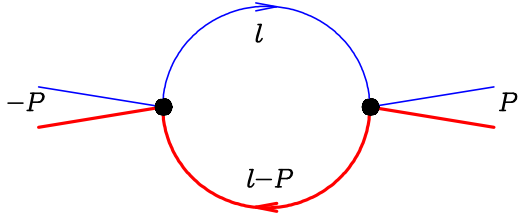}
\end{center}
\caption{Diagram for the calculation of the coefficient of the
scalar integral $B_0(P;0,m)$. The massive quark is 
represented by the heavy (red) line. The momenta on the external lines are all outgoing.}
\label{bubble}
\end{figure}

To set up the formalism 
we consider the scalar bubble integral with one massive propagator, as shown in Fig.~\ref{bubble}
(bubbles 4 and 5 are of this type).
The evaluation of the coefficient of this bubble integral starts from the identity,
\beq
\int d^4l \, \delta^{(+)}(l^2) \delta^{(+)}((l-P)^2-m^2)
 = \int_0^\infty t dt \int \spa l.{dl} \spb l.{dl} \delta(P^2-m^2-t\spab l.P.l) \;,
\eeq
so that the formalism follows through as in the massless case, except that $t$ is now frozen at the value,
\beq
t = \frac{P^2-m^2}{\spab l.P.l} \;.
\eeq
The discontinuity of the scalar integral is given by,
\beqn
\Delta I_2(P,m) &=& \int d^4l \, \delta^{(+)}(l^2) \delta^{(+)}((l-P)^2-m^2) \nn \\
&=& \int \frac{\spa l.{dl} \spb l.{dl}}{\spab l.P.l} \left( \frac{P^2-m^2}{\spab l.P.l} \right) \nn \\
 &=& (P^2-m^2) \int \frac{\spa l.{dl} \spb\eta.l}{\spab l.P.\eta \spab l.P.l } \;.
\eeqn
Using the standard formula~\cite{Cachazo:2004kj,Britto:2005ha} 
we can express the integrand as a total derivative, 
and perform one of the spinor integrations,
\beq
\label{total_derivative_id}
\spb{l}.{dl}
\left(\frac{\spb{\eta}.{l}^n}{\langle l|P|l]^{n+2}}\right)
= \spb{dl}.{\partial_l} \left(\frac{1}{n+1}
\frac{1}{\langle l|P|\eta]}
\frac{\spb{\eta}.{l}^{n+1}}{\langle l|P|l]^{n+1}}\right) \; .
\eeq
We obtain,
\beq
\Delta I_2(P,m) = (P^2-m^2) \int \frac{\spa l.{dl} \spb\eta.l}{\spab l.P.\eta \spab l.P.l } \;,
\eeq
where $\eta$ is an arbitrary massless momentum, 
We perform the final integral over $\spa l.{dl}$ by inspecting the 
residues of the pole in the integrand when $\braa l = \brab\eta P$ to find,
\beqn
\Delta I_2(P,m) &=&  - (P^2-m^2) \frac{\spba\eta.P.\eta}{P^2 \spab\eta.P.\eta} \\
&=& -\left(\frac{P^2-m^2}{P^2}\right) \;.
\eeqn
So, in contrast to the massless case, when applying the spinor integration approach an additional rescaling
factor of $P^2/(P^2-m^2)$ must be applied in order to obtain the coefficient of the scalar bubble integral.

\subsection{Calculation of the rational terms}

\label{rational}
The purely rational terms are calculated using a traditional Feynman
diagram approach, with tensor integrals handled using
Passarino-Veltman reduction~\cite{Passarino:1978jh}. This might seem
like a retrograde step, ineluctably leading to the algebraic
complexity that we have been trying to avoid in this amplitude
calculation.  However two details of our particular case make
it quite simple.  

\begin{figure}
\begin{center}
\includegraphics[angle=270,scale=0.5]{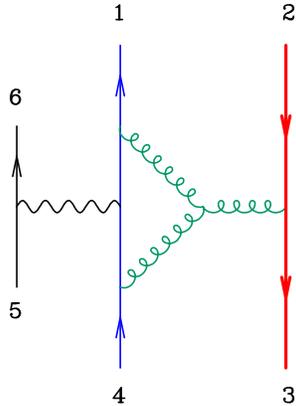}
\end{center}
\caption{The box diagram contributing to the rational part in leading colour.}
\label{ratbox}
\end{figure}

First, in the calculation of the rational part 
we only need to consider diagrams that
violate the cut-constructibility condition~\cite{Bern:1994cg}. This
states that if $n$-point integrals ($n>2$) have at most $n-2$ powers of the
loop momentum in the numerator of the integrand, and the two-point
integrals have at most one power of the loop momentum, they
will be cut-constructible. For our particular calculation the pentagon
diagrams are all cut-constructible, and, in leading colour, there is
only one box diagram, shown in Fig.~\ref{ratbox},
which is not cut-constructible and hence gives
a rational part.  The calculation of the rational parts from lower point diagrams,
$n=3,2$ does not lead to great algebraic complexity.
Second, the rational terms are
generated by terms in the Passarino-Veltman decomposition that
involve the metric tensor, $g^{\mu \nu}$.  Therefore any terms that
are not proportional to $g^{\mu\nu}$ may be discarded at an early
stage of the computation. 

The results for the rational terms turn out to be quite simple. For the $\{-,+\}$ case we find,
\beqn
&&R^\lc(1_\q^-,2_\Qb^+, 3_\Q^-,4_\qb^+,5_\ellb^+,6_\ell^-) =
  -\frac{2}{9} (-i) \Atree_6(1_\q^-,2_\Qb^+, 3_\Q^-,4_\qb^+,5_\ellb^+,6_\ell^-) \nn \\
&+& \Bigg\{ \frac{\spa1.6}{2 s_{56}} \, \Bigg[ 
   \frac{ \spa {1}.{3} \spb {1}.{5} \spb {1}.{2} }{\spab {1}.{(2+3)}.{1}  \spab {4}.{(2+3)}.{1} }
 - \frac{\spa1.3 \spb1.5}{s_{23} \spab {1}.{(2+3+4)}.{1}} \Big(
   \frac{ s_{234} \spb {1}.{2}}{  \spab {4}.{(2+3)}.{1} } + \spb {2}.{4} \Big)
\nn \\
&+&\frac{\spb2.4}{s_{234}} \Big(
   \frac{ \spb {2}.{5} }{\spb {2}.{3}} 
 - \frac{ \spa {3}.{4} \spb {4}.{5} }{\spab {4}.{(2+3)}.{4}}
   \Big) \Bigg]\Bigg\}
 -\Bigg\{{\rm flip }\Bigg\} \;,
\eeqn
where the flip operation is defined in Eq.~(\ref{flip1def}).
The contribution for the opposite spin labels on the heavy quark line
is simply related to this one,
\beq
  R^\lc(1_\q^-,2_\Qb^-, 3_\Q^+,4_\qb^+,5_\ellb^+,6_\ell^-)
= R^\lc(1_\q^-,3_\Qb^+, 2_\Q^-,4_\qb^+,5_\ellb^+,6_\ell^-) \;.
\eeq
For the $\{-,-\}$ spin labels the result is,
\beqn
\label{Rmm}
&&R^\lc(1_\q^-,2_\Qb^-,3_\Q^-,4_\qb^+,5_\ellb^+,6_\ell^-) =
  -\frac{2}{9} (-i) \Atree_6(1_\q^-,2_\Qb^-,3_\Q^-,4_\qb^+,5_\ellb^+,6_\ell^-) \nn \\
&+& {\cal N}_{--} \Bigg( \Biggl\{
    \frac{\spa1.6}{2 s_{56}} \, \Biggl[
    \frac{\spb1.5}{\spab 1.{2+3+4}.1 s_{23}} \Big(  
    \frac{s_{234} (\spab 1.2.1 -\spab 1.3.1)}{\spab 4.{2+3}.1}
   - (\spab 1.2.4 -\spab 1.3.4) \Big) \\
&-& \frac{(\spab1.2.1-\spab 1.3.1)}{\spab 1.{2+3}.1} \Big(  
    \frac{\spb 1.5}{\spab 4.{2+3}.1} 
  + \frac{\spb 4.5}{s_{23}} \Big) 
 \Biggr]\Bigg\} +\Bigg\{{\rm flip} \Bigg\}\Bigg)\;, \nn
\eeqn
where the $\mbox{flip}$ symmetry applies to all terms inside the curly brackets $\{ \ldots \}$ but not the
prefactor ${\cal N}_{--}$, defined in Eq.~(\ref{normalization}). The final combination of spin labels is obtained by
symmetry,
\beq
R^\lc(1_\q^-,2_\Qb^+,3_\Q^+,4_\qb^+,5_\ellb^+,6_\ell^-)
 = -\mbox{flip} \left( R^\lc(1_\q^-,2_\Qb^-,3_\Q^-,4_\qb^+,5_\ellb^+,6_\ell^-) \right) \;,
\eeq
which we note is equivalent to replacing ${\cal N}_{--}$ by ${\cal N}_{++}$ in Eq.~(\ref{Rmm}).

\section{The results for primitive amplitude $\Asl_6$}

The subleading colour primitive amplitude is shown in Fig.~\ref{Asl}.
Fortunately with our choice of massive spinors,
Eq.~(\ref{Massivespinordefs}), the entire result for all spin labels
of the massive quark line can be obtained from the one-loop results 
of Bern, Dixon and Kosower in Eqs.~(12.10,12.11) of ref.~\cite{Bern:1997sc}, 
after some replacements and manipulation.  The calculation reported in
ref.~\cite{Bern:1997sc} was for zero mass quarks and leptons with the
following helicity assignments,
\beq \label{BDKnotation}
\Asl_6(1^+_\Q, 2^-_\Qb, 3^+_\q, 4^-_\qb,5^-_\lb,6^+_\l)\; .  
\eeq
According to Fig.~\ref{lowestorder} our standard labelling of the
graphs (for one of the amplitudes whose spin labelling 
corresponds to a non-zero amplitude in the massless case) is,
\beq
\label{ournotation}
\Asl_6(1^-_\q, 2^+_\Qb, 3^-_\Q, 4^+_\qb,5^+_\lb,6^-_\l) \;.
\eeq

To establish the correspondence between our massive amplitudes and 
the massless ones of ref.~\cite{Bern:1997sc},
we first consider the tree graphs. In our notation the tree amplitudes are given by 
Eqs.~(\ref{Treehelicityconserving},\ref{Treehelicityviolating}).
Note that we must perform the interchanges $(1 \leftrightarrow 3), \langle \rangle \leftrightarrow []$),
to compare with Eq.~(12.9) of ref.~\cite{Bern:1997sc}. As a consequence of our 
Eq.~(\ref{helicityconserving}) 
the massive $\{-,+\}$ tree-level amplitude, Eq.~(\ref{Treehelicityconserving}),
calculated from diagrams of Fig.~\ref{lowestorder} 
and expressed in terms of the massless vectors, $k_i$,
is identical to the massless result presented in Eq.~(12.9) of ref.~\cite{Bern:1997sc}
after performing the above interchange.

This same transformation, flipping the sign of all helicities
and interchanging 1 and 3, can be used at one-loop level to obtain 
the bulk of the results for the massive theory. This is true
for the upper two diagrams of Fig.~\ref{Asl}, because the massive quark
enters only in the form of the heavy quark current see, Eqs.~(\ref{helicityconserving},\ref{helicityviolating}).
The primitive amplitude shown in Fig.~\ref{Asl} can be split into 
divergent ($V^{\sl}$) and finite ($F^{\sl}$) pieces as follows
\begin{equation}
\label{VFdecompSL}
\Asl_6 = \Bigl[ \Atree_6 V^{\sl} + i \, F^{\sl} \Bigr]\ ,
\end{equation}
After performing the interchanges to reduce it to our notation,
the singular part of the one-loop amplitude as reported by Bern, Dixon and Kosower in ref.~\cite{Bern:1997sc},
Eq.~(12.10) is, 
\beq
V^{\sl}(1,2,3,4) = V^{\rm box}(1,2,3,4)  + V^{\rm vertex}(1,2,3,4)  
\eeq
where the two contributions correspond to the upper and lower row of Fig.~\ref{Asl} in the massless theory,
\beqn
V^{\rm box}(1,2,3,4) &=& 
\biggl[- \frac{1}{\e^2} \left(\frac{\mu^2}{-s_{14}}\right)^\e 
  - \frac{3}{2\e} \left(\frac{\mu^2}{-s_{14}}\right)^\e - 4  \biggr] \;, \\
V^{\rm vertex}
(1,2,3,4) &=& 
  \biggl[ - \frac{1}{\e^2} \left(\frac{\mu^2}{-s_{23}}\right)^\e 
  - \frac{3}{2 \e} \left(\frac{\mu^2}{-s_{23}}\right)^\e 
  - \frac{7}{2} \;,
\biggr] 
\label{masslessvertex}
\eeqn
where $s_{ij} =2 k_i \cdot k_j$. $V^{\rm vertex}$ in Eq.~(\ref{masslessvertex})
is the complete correction to external 
vertex for a massless line in the FDH scheme.
For the massive case this must be replaced by the vertex correction 
for a massive line, i.e.\ the result for the 
lower two diagrams of Fig.~\ref{Asl}. The result is,
\beqn
V^{\rm vertex}_{+ \, -} =
V^{\rm vertex}_{- \, +} &=& \Big(\frac{\mu^2}{m^2}\Big)^\e \Big( \frac{1}{2 \e} 
-\frac{1}{2 \beta} \big(1+\beta^2 \big) 
 \Big[\frac{\ln x}{\e}+G(x)\Big] -\frac{3}{2} \beta \ln x+\frac{1}{2}\Big) \;, \\
V^{\rm vertex}_{+ \, +} =
V^{\rm vertex}_{- \, -} &=& V^{\rm vertex}_{- \, +} + \frac{1}{2} \beta \ln x \;,
\eeqn
where $\beta$ and $\beta_\pm$ are given in Eq.~(\ref{betadef}), $x=-\beta_-/\beta_+$, and,
\beq
G(x) = -2~ \li(-x) -2 \ln x \ln(1+x)+ \frac{1}{2}\ln^2(x)-\frac{\pi^2}{6},\;\;\; 
\li(x) = - \int_0^x \, \frac{dz}{z} \, \ln(1-z) \; .
\eeq
%
The self energy corrections on the external massive lines will be accounted for 
separately in association with the wave function renormalization.
This concludes our description of the divergent parts 
and the lower two graphs of Fig.~\ref{Asl}. 

We now turn to the finite parts of the massive primitive amplitudes shown
in the upper part of Fig.~\ref{Asl}. 
Because of Eq.~(\ref{helicityconserving}), the results for the 
$\{+,-\}$ and $\{-,+ \}$ amplitudes from the upper row of Fig.~\ref{Asl}
in the {\it  massive} theory, expressed in terms of the lightlike momenta $k_i$,
are given by the results in the {\it massless} theory. 

The only remaining issue is whether we can also obtain the massive 
$\{-,-\}$ and $\{+,+\}$ amplitudes from the massless results.
Note that in ref.~\cite{Bern:1997sc} the finite parts of the massless 
amplitudes are written in terms of
certain symmetry operations in order to make the amplitudes more compact.
As a first step we write out the amplitudes explicitly. With this result in 
hand we want to address the issue of whether the results obtained with an external
fermionic current, Eq.~(\ref{helicityconserving}) (i.e.\ the massless one-loop amplitude), 
can be used to obtain the results with the fermionic currents of 
Eq.~(\ref{helicityviolating}). Thus expressed in our notation, the massless 
amplitude for helicity choice ($2^+_\Qb,3^-_\Q$) must contain
one $\langle 3|$ and one $|2]$. All other dependence on $k_2$ or $k_3$
can only enter in the combination
$k_2+k_3$ which can be eliminated by momentum conservation. After the amplitude
has been recast in this form, we can obtain the required result by replacing the current 
of Eq.~(\ref{helicityconserving}) with the current of Eq.~(\ref{helicityviolating}).

An example may help to clarify the procedure. We shall consider a particular box which contributes to 
the one-loop amplitude,
\beqn
&&\Asl_6(
1_\q^-, 2_\Qb^{h_2},3_Q^{h_3}, 4_\qb^+,5_\ellb^+,6_\ell^-)=d_{4|1|23}(h_3,h_2) \; D_0(k_4,k_1,p_{23};0,0,0,0) + \ldots
\eeqn
The result for this box coefficient in the massless helicity-conserving amplitude,
(adapted from the first line of Eq.~(12.11) of ref.~\cite{Bern:1997sc}) is,
\beqn
&&d_{4|1|23} (-,+)= s_{14} \Bigg( \frac{ {\spa1.3}^2 {\spb4.5}^2 }{\spa3.2 \spb5.6  \spba4.{(2+3)}.1}
- \frac{ {\spba1.{(2+3)}.6}^2 {\spba 2.{(1+3)}.4}^2}
 {\spb3.2 \spa5.6 {\spba 1.{(2+3)}.4}^3}  \Bigg) \; .
\eeqn
This can be rewritten in a form which makes the $|3 \rangle$, $|2 ]$ and $k_2+k_3$ 
structure manifest,
\beqn
&&d_{4|1|23} (-,+)=-s_{14} \nn \\
&&\times \Bigg( \frac{\spa1.3 \spab1.{(2+3)}.2  {\spb4.5}^2 }
{s_{23} \spb5.6 \spba4.{(2+3)}.1}
- \frac{ {\spba1.{(2+3)}.6}^2 \spaa3.{(2+3)(1+2+3)}.4  
 \spba 2.{(1+2+3)}.4}
 {s_{23} \spa5.6 {\spba1.{(2+3)}.4}^3 }  \Bigg) \;. \nonumber \\
\eeqn

Replacing the $\{ -,+\}$-current, Eq.~(\ref{leftright}), with the $\{-,-\}$-current, 
Eq.~(\ref{leftleft}), 
we obtain the result for the coefficient of this box in the massive amplitude labelled $\{-,-\}$
\beqn
&&d_{4|1|23} (-,-)=-{\cal N}_{--}\; s_{14} \nn \\
&&
\times \Bigg(\frac{\spa1.2 \spa1.3 {\spb4.5}^2 }
{\spa3.2 \spb5.6 \spba4.{(2+3)}.1}
+ \frac{ {\spba1.{(2+3)}.6}^2{\spba2.{(1+3)}.4}{\spba3.{(1+2)}.4}}
{ \spb3.2 \spa5.6 {\spba1.{(2+3)}.4}^3} \Bigg) \;.
\eeqn
Carrying out this operation for all the terms in Eq.~(12.11) of ref.~\cite{Bern:1997sc}
we obtain the complete massive amplitude for the $\{-,-\}$ and $\{+,+\}$ spin labellings.

\section{Results for the primitive amplitudes $A_6^{lf}$ and $A_6^{hf}$.}
\begin{figure}
\begin{center}
\includegraphics[angle=270,scale=0.4]{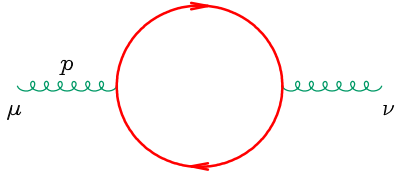} 
\end{center}
\caption{Fermion loop contribution to the gluon vacuum polarization.
\label{fermionloop}}
\end{figure}

The unrenormalized contribution of the fermion loop diagram,
shown in Fig.~\ref{fermionloop}, for a quark of mass $m$ is, 
\beq
\Pi^{\mu \nu}  = i g^2 \cg \Big[ g^{\mu \nu} p^2 -p^\mu p^\nu \Big] \Pi(p^2,m^2) \; ,
\eeq
where (not including the minus sign for a fermion loop),
\beq
\label{Pidef}
\Pi(p^2,m^2)  = 2 T_R \; \frac{1}{\e} \Big(\frac{\mu^2}{m^2}\Big)^\e \; 
\int_0^1 \; dx \frac{4 x(1-x)}{\Big[1-x (1-x) \frac{p^2}{m^2}\Big]^\e} \;,
\eeq
and $T_R=\frac{1}{2}$.
For $n_{lf}$ quarks which can be considered massless this becomes,
\beq
\Pi(p^2,0)= n_{lf} 
\Big[\frac{2}{3}\frac{1}{\e}  \Big(\frac{\mu^2}{-p^2}\Big)^\e +\frac{10}{9}\Big]+O(\e) \;. 
\eeq
Thus in our notation, (c.f.\ Eq.~(\ref{Oneloopcolourdecomp})) the result for the unrenormalized
fermion-loop primitive is,
\beq
A_6^{lf}(1_\q^+,2_\Qb^\pm,3_\Q^\mp,4_\qb;5_\ellb^+,6_\ell^-
)
 = \Atree_6(1_\q^+,2_\Qb^\pm,3_\Q^\mp,4_\qb;5_\ellb^+,6_\ell^-)
  \, \left[ \frac{2}{3} \frac{1}{\e} \left( \frac{\mu^2}{-s_{23}}\right) ^{\e}
          + \frac{10}{9} \right] \,,
\eeq
where $A_6^{\rm tree}$ are given in Eqs.(\ref{Treehelicityconserving},~\ref{Treehelicityviolating}).

\section{Renormalization}
The amplitudes presented so far are bare amplitudes, which require
ultraviolet renormalization. The renormalization scheme is slightly
more complicated in the presence of massive particles~\cite{Collins:1978wz}
so we specify it in detail here.
The requirements for our renormalization scheme are,
\begin{itemize}
\item
The decoupling of heavy quarks should be manifest.
\item The evolution equations for the running coupling and for the 
parton distribution functions should be the same as the equations 
in the theory without the heavy quark. Both the strong coupling 
and the parton distribution functions should run with the 
coefficients appropriate for the $\overline{\rm MS}$ scheme in the absence of the massive particles.
\item The mass parameter should correspond to a pole mass.
\end{itemize}
These three requirements completely specify the renormalization scheme.
If the diagram in question contains no heavy internal loops of heavy particles
we use the $\overline{\rm MS}$ scheme. If on the other hand the diagram contains heavy loops
we will perform subtraction at zero momentum, $p=0$. 
The resultant renormalized Green's function 
will be a function of $p^2/m^2$ and hence exhibit decoupling as the mass of the heavy quarks becomes large.

The full renormalized amplitude, $A_{6;1}^R$
is obtained by adding an overall counterterm, 
\beqn \label{overall}
&&N_c A_{6;1}^R = N_c A_{6;1} \nn \\
&+& g^2 \cg \Big\{ -2 \Big(\frac{11}{6} N_c-\frac{n_{\lf}}{3}\Big)\frac{1}{\e}
+\frac{2 n_{\hf}}{3}\Big(\frac{1}{\e}  +\ln\frac{\mu^2}{m^2}\Big)
+ \frac{N_c}{3} 
-C_F \Big(\frac{3}{\e}  +3 \ln\frac{\mu^2}{m^2}+5\Big)
\Big\}\Atree_6 \; .\nn \\
\eeqn
We will now justify the contributions in the counterterm term by term.
The first term in Eq.~(\ref{overall}) 
is the normal $\overline{MS}$ renormalization
of the coupling constant, which includes the renormalization of the $n_\lf$ loops of massless fermions.
The second term in Eq.~(\ref{overall}) deals with the case where we have $n_\hf$ heavy fermions.
Renormalizing Eq.~(\ref{Pidef}) at $p^2=0$ to ensure decoupling of the heavy fermions as $m$ becomes large,
we obtain for each of the $n_{\hf}$ heavy fermions,
\beq
\Pi^R(p^2,m^2)=-4 \int_0^1 \; dx \; x(1-x) \; \ln\Big(1-x(1-x)\frac{p^2}{m^2}\Big) \;.
\eeq
In the high mass limit $\Pi^R$ simplifies to,
\beq
\Pi^{R}(s,m^2) \to  \frac{2}{15} \frac{s}{m^2}+O(\frac{s^2}{m^4}) \;.
\eeq
Thus after renormalization the contribution of the heavy quark 
is given by (c.f.\ Eq.~(\ref{Oneloopcolourdecomp})) 
\beq
A_6^{hf}(1_\q^+,2_\Qb^\pm,3_\Q^\mp,4_\qb;5_\ellb^+,6_\ell^-)
 = \Atree_6(1_\q^+,2_\Qb^\pm,3_\Q^\mp,4_\qb;5_\ellb^+,6_\ell^-)\Pi^R(s_{23},m^2) \;.
\eeq
We must also perform 
a finite renormalization 
of the gauge coupling~\cite{Kunszt:1993sd} 
to translate from the FDH coupling to the normal ${\overline{\rm MS}}$ coupling,
\begin{equation}
\alpha_s^{{\rm FDH}}=\alpha_s^{\overline{\rm MS}}
\left(1+ \frac{N_c}{6} \frac{\alpha_s^{\overline{\rm MS}}}{2\pi}\right) \; .
\end{equation}
This explains the third term in Eq.~(\ref{overall}).
The last term in Eq.~(\ref{overall}) represents the wave function
renormalization for the two external massive fermions, calculated in Appendix~\ref{fermionse}.
In the FDH scheme we have from Eq.~(\ref{wavefunction}), 
\beq Z_Q= 1-g^2 \cg C_F \Bigg[\frac{3}{\e}+3
\ln\left(\frac{\mu^2}{m^2}\right)+5\Bigg]+O(g^4,\e)\; , 
\eeq
independent of the gauge-fixing parameter in any covariant gauge.

Since our calculation is performed in the four-dimensional helicity scheme
there is a further finite renormalization~\cite{Kunszt:1993sd} required to arrive at the 
't Hooft-Veltman scheme. We shall work consistently in the FDH scheme,
so this will not be required.

\section{Implementation into MCFM}
\label{MCFMsec}
\begin{figure}
\begin{center}
\includegraphics[angle=270,scale=0.6]{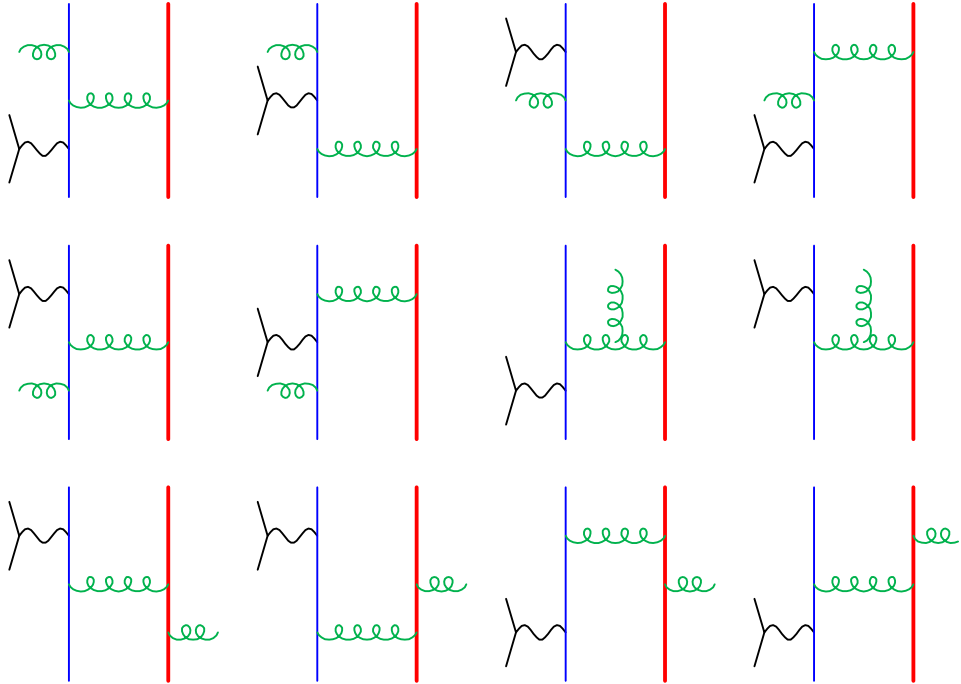} 
\end{center}
\caption{Feynman graphs appearing in the calculation of the real radiation amplitudes.
\label{real}}
\end{figure}

The one-loop matrix elements, computed using the methods described above, have been included
in a full next-to-leading order calculation of the $WQ\bar Q$ process. 
The amplitude for the lowest order process is given in 
Eqs.~(\ref{Treehelicityconserving},\ref{Treehelicityviolating}).
In order to complete the NLO calculation the Born level amplitude and the one-loop amplitude 
must be supplemented with results for the real radiation diagrams
and a method for cancelling infrared singularities between the two contributions. The
tree-level real radiation process has been computed using the diagrams shown in
Fig.~\ref{real}, adopting the same choice of massive spinors as used in the
virtual contribution.
Infrared singularities are handled using the subtraction method~\cite{Ellis:1980wv}
implemented using the dipole formulation~\cite{Catani:1996vz} and 
extended to the case of massive emitters and
spectators~\cite{Catani:2002hc}. The full calculation
will be made available as part of the the MCFM code~\cite{MCFM,Campbell:1999ah}.

To provide a point of comparison we have timed the
evaluation of the complete one-loop contribution, i.e. the interference 
in Eq.~(\ref{looptreeint}), summed over all four possible spin labels on the heavy quark line.
Using a standard $2.66$~GHz machine and compiling the code with the {\tt -O2} flag of {\tt gfortran}
this operation takes $4$ms for a single phase space point, which is about a factor of sixty slower
than evaluating the corresponding massless amplitudes as implemented in MCFM.

\subsection{Checks on the calculation}
A number of checks have been performed at various stages of the calculation.
\begin{enumerate}
\item At the level of the one-loop amplitude, we have cross-checked our calculation with a
numerical implementation of generalized $d$-dimensional unitarity~\cite{Ellis:2008qc}. 
This check confirms the
coefficients of all the scalar integrals and rational terms, as well as the complete
(unrenormalized) amplitude, at any given phase space point.
The values of the unrenormalized one-loop amplitudes at a specific phase space point
are given in Appendix~\ref{numer}. These numbers may be useful in performing 
a check of our calculation.
\item The cancellation of infrared singularities is performed using a slight extension of
the original dipole formulation in which the extent of each subtraction region is controlled
by an additional parameter~\cite{Nagy:1998bb,Nagy:2003tz}. This parameter also appears in the
integrated form of the dipole counterterms in such a way that the sum of real and virtual
radiation does not depend upon its value. We have checked that this independence is indeed
manifest in our calculation.
\item We have checked that our final results for the $Wb\bar b$ integrated cross section
agree with the values reported in the earlier calculations of
Refs.~\cite{Febres Cordero:2006sj,Cordero:2009kv,Cordero:2010qn}.
\end{enumerate}

\subsection{Phenomenology}
\begin{table}
\begin{tabular}{l@{\hspace{1cm}}l}
$m_W = 80.44$~GeV & $m_b=4.62$~GeV \\ 
$m_t = 172.6$~GeV & $\sin^2\theta_w = 0.223$ \\
$G_F = 1.16639 \cdot 10^{-5}~\mbox{GeV}^{-2}$
 & $\alpha = \frac{\sqrt{2}}{\pi} G_F m_W^2 \sin^2\theta_w $ \\ 
$V_{ud}=V_{cs} = 0.974$~GeV & $V_{us}=V_{cd} = 0.227$ \\
\begin{tabular}{l}
 PDF set: CTEQ6L1 \\ 
 $\alpha_s(m_W+2m_b)=0.130345$
 \end{tabular} $\Biggr\}$ LO &
\begin{tabular}{l}
 PDF set: CTEQ6M \\ 
 $\alpha_s(m_W+2m_b)=0.118298$
 \end{tabular} $\Biggr\}$ NLO \\
\end{tabular}
\caption{Inputs used for the results presented in section~\ref{MCFMsec}.
\label{params}}
\end{table}
For now we present only a limited set of results, focussing mainly on the comparison with
previous calculations and presenting a simple distribution involving the decay products 
of the $W$ boson. We leave a detailed phenomenological
study for a future work.

For convenience, we choose the same set of input parameters here as reported in
Ref.~\cite{Cordero:2009kv}, which are summarized in Table~\ref{params}.
The final state is defined by the following cuts on the $b$-jets,
\beq
p_T^b > 25~\mbox{GeV} \;, \qquad
|\eta^b| < 2.5 \;,
\eeq
where the jets are identified using the $k_T$ clustering algorithm with pseudo-cone
size $R=0.7$. The results presented here are inclusive of the additional jet
that may be present at NLO. 

As already discussed previously, although we
treat the $b$-quark as a massive particle when it appears in the final
state, we use $n_{\lf}=5$ light flavours in the running of $\alpha_s$ and
the PDF evolution. This is primarily for comparison with previous work~\cite{Cordero:2009kv}.
We note that because of the smallness of $V_{cb}$ and $V_{ub}$ the $b$-quark distributions
in the initial state make a negligible contribution. However because $s_{23}> 4 m_b^2$ it is
more appropriate to have a strong coupling constant running with $n_{\lf}=5$ active flavours.

We first present the cross sections for this process at the LHC, for center-of-mass
energies $\sqrt{s}= 7$, $8$ and $14$~TeV. The results at LO and NLO are shown in
Table~\ref{results} where we have used a scale choice $\mu_R = \mu_F = m_W + 2m_b$
throughout, again to facilitate comparison with Ref.~\cite{Cordero:2009kv}.
For these results no cuts are applied on the decay products of the $W$ boson and the corresponding
branching ratio is removed, so that the reported cross sections are for a $W$ that does not decay. In this
way, one sees from the final column ($\sqrt{s}=14$~TeV) that our results agree with those reported
in Ref.~\cite{Cordero:2009kv} at the level of $0.5$\%. As noted in earlier work~\cite{Campbell:2003hd},
at LHC energies the NLO corrections
are substantial because of the influence of the quark-gluon initial state.
\begin{table}
\begin{tabular}{|c@{\hspace{0.5cm}}l@{\hspace{0.5cm}}|c|c|c|}
\hline
\multicolumn{2}{|c|}{$\sqrt{s}$} & ~~~$7$~TeV~~~ & ~~~$8$~TeV~~~ & ~~~$14$~TeV~~~ \\ \hline
\multirow{2}{*}{$W^+b\bar b$} &  LO & 4.456(2)  & 5.157(2) & 9.041(3) \\ 
                              & NLO & 8.655(9)  & 10.58(2) & 23.51(3) \\ \hline
\multirow{2}{*}{$W^-b\bar b$} &  LO & 2.588(2)  & 3.109(1) & 6.256(2) \\ 
                              & NLO & 5.053(5)  & 6.353(6) & 15.55(2) \\ \hline
\end{tabular}
\caption{LO and NLO cross sections (in picobarns) for $Wb\bar b$ production at various energies of the LHC.
Integration errors are shown in parentheses.
\label{results}}
\end{table}

\begin{figure}
\begin{center}
\includegraphics[angle=0,scale=0.8]{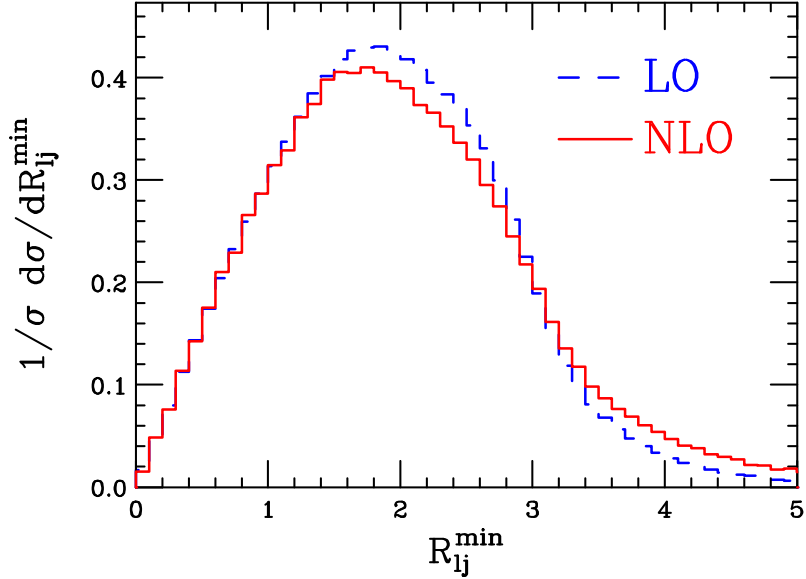}
\end{center}
\caption{Distribution of the minimum separation between the electron and any jet, $R_{lj}$, defined
in Eq.~(\ref{Rlj}) of the text, for $W^-b{\bar b}$ production at the $14$~TeV LHC. The distributions
are normalized to unit area.
\label{Rljfig}}
\end{figure}

To illustrate a new quantity that may now be computed at NLO accuracy, in Fig.~\ref{Rljfig} we show
the LO and NLO predictions for the quantity $R_{lj}^{\rm{min}}$ which is defined as the separation 
between the charged lepton and the closest jet,
\beq
R_{lj}^{\rm{min}} = \min_{\{\rm{jets}\}} \sqrt{
 (\eta_{\rm{lepton}} - \eta_{\rm{jet}})^2 + (\phi_{\rm{lepton}} - \phi_{\rm{jet}})^2 } \;.
\label{Rlj}
\eeq
In this equation the azimuthal angles and pseudorapidities (in the lab-frame) of the lepton and
jets are denoted by $\phi$ and $\eta$ respectively.
We see that the effect of the NLO corrections on the shape of this
distribution is relatively minor, visible only for
$R_{jl}^{\rm min} > 1.5$.
\section{Conclusions}

We have presented the first computation of the $W b \bar{b} $ cross
section with massive $b$-quarks, including the lepton correlations
present in the decay of the $W$-boson.  This calculation required
knowledge of the one-loop virtual corrections to the $q \Qb \Q \qb
\ellb \ell$ process retaining the mass for the heavy quarks $Q$. The
one-loop amplitude was obtained using the spinor helicity
formalism. This method has been extensively used for one-loop
calculations with massless quarks, but rarely with massive quarks.
The calculation required a number of modifications of standard
techniques to cope with the presence of the mass.

Although our results are analytic we have not yet simplified them
sufficiently to publish them in a journal article.  Our results for the
one-loop amplitudes will
be included in the released version of MCFM, which is an appropriate
method of publishing such results. Our analytic results 
did lead to a code which is
fast and numerically stable. Using this code we intend to study the
detailed phenomenology of this process in a future publication.

\section*{Acknowledgements}
We are happy to acknowledge useful discussions with Fabrizio Caola and Kirill 
Melnikov.  The work of SB has been supported 
in part by Danish Natural Science Research Council grant 10-084954.
Fermilab is operated by Fermi Research Alliance, LLC, 
under contract DE-AC02-07CH11359 with the United States Department of Energy.
\appendix
\section{Notation for spinor products}
\label{Notation}
The spinor notation is almost standard, but because of the presence of 
massive particles, it is important to make it explicit.
We adopt the following notation for massless spinors,
\begin{eqnarray}
|i\rangle &= |i+\rangle = u_+(k_i), \; |i] &= |i-\rangle = u_-(k_i) \;, \nn \\
\langle i| &= \langle i-| = \bar{u}_-(k_i),\; [i| &= \langle i+| = \bar{u}_+(k_i)  \;.
\end{eqnarray}
Further the spinor products are defined as,
\begin{eqnarray}
\spa i.j &=& \langle i-|j+\rangle = \bar{u}_-(k_i) u_+(k_j) \;, \nn \\
\spb i.j &=& \langle i+|j-\rangle = \bar{u}_+(k_i) u_-(k_j) \;,
\end{eqnarray}
with $k_i,k_j$ massless particles. With our convention,
\begin{equation}
\spa i.j \; \spb j.i = 2 k_i \cdot k_j = s_{ij} \;.
\end{equation}
We also define the spinor strings.
\begin{eqnarray}
\AB {i}{j}{k}  &\equiv &  \langle k_i -| \slsh{k}_j | k_k - \rangle \;, \nn \\
\AB {i}{j+k}{l} &\equiv & \langle k_i -| (\slsh{k}_j+\slsh{k}_k) | k_l - \rangle \;, \nn \\
\spaa {i}.{jk}.{l} &\equiv & 
\langle k_i -| \slsh{k}_j \slsh{k}_k  | k_l + \rangle \;. 
\end{eqnarray}
For the case of a massless momentum $k_j$ we may write,
\begin{equation}
\AB {i}{j}{k} =  \spa i.j \spb j.k \;,
\end{equation}
but for the case of a massive momentum, (in our notation the momenta $p_2$ and $p_3$),
this separation in no longer possible. As a compact notation we therefore write in 
this case $\AB {i}{\slsh{{\bf j}}}{k}$, denoting the massive momentum by a 
bold-face symbol.

The Schouten identity,
\begin{equation}
\label{Schouten}
[  i \, k ] [  m \, n ] =[  i \, n ] [  m \, k ]+ [  i \, m ] [  k \, n ] \;,
\end{equation}
may be applied to these compound quantities. Thus we have,
\beq
\AB {i}{\slsh{{\bf j}}}{k} [m \, n] =
 \AB {i}{\slsh{{\bf j}}}{n} [m \, k] 
+ \AB {i}{\slsh{{\bf j}}}{m} [k \, n] \;.
\eeq
\section{Tree level results}
\label{treeapp}
\subsection{Results for $A(1_g,2_\Qb,3_Q,4_g)$}
\label{gQbQg}
As an example of the use of the massive spinors employed in the calculation we consider the tree-level two-quark two-gluon amplitude,
$A(1_g,2_\Qb,3_Q,4_g)$. The fermions, with momenta $p_2$ and $p_3$ have a common mass $m$.
The momenta are all outgoing so that $k_1+p_2+p_3+k_4=0$.
\begin{figure}
\begin{center}
\includegraphics[angle=270,scale=0.5]{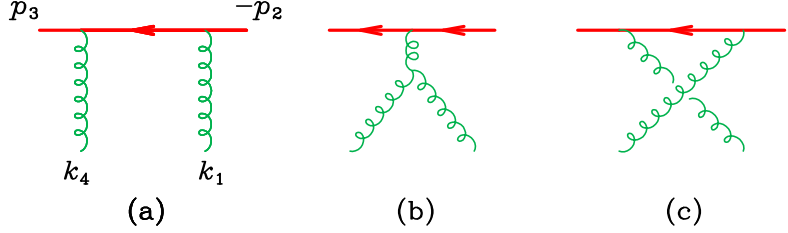}
\end{center}
\caption{Diagrams for two gluons coupling to a massive quark line.}
\label{qqgg}
\end{figure}
The amplitude is obtained by summing the contributions of the three Feynman diagrams,
$M^{(a)}$, $M^{(b)}$ and $M^{(c)}$ shown in Fig.~\ref{qqgg},
\beqn
A_4^{\rm tree} = -\frac{i}{2} g^2 \Bigg[ (T^{C_4} T^{C_1})_{i_3 i_2} (M^{(a)}+M^{(b)})
 +(T^{C_1} T^{C_4})_{i_3 i_2} (M^{(c)}-M^{(b)}) \Bigg] \;,
\eeqn
Normalizing the colour matrices as follows,
\beq
{\rm Tr}~ T^A T^B =\delta^{AB} \;,
\eeq
an explicit calculation yields the following contributions,
\beqn
M^{(a)}
 &=& \bar{u}(p_3) \slsh{\e}_4 \frac{\slsh{p_{34}}+m}{p_{34}^2-m^2} \slsh{\e}_1 v(p_2) \;, \\
M^{(b)} &=& 
\frac{\e_1\cdot \e_4}{k_1\cdot k_4}
 \bar{u}(p_3) \slsh{k}_1v (p_2) 
-\frac{k_1\cdot \e_4}{k_1\cdot k_4}
 \bar{u}(p_3) \slsh{\e}_1 v (p_2) 
+\frac{k_4\cdot \e_1}{k_1\cdot k_4}
 \bar{u}(p_3) \slsh{\e}_4 v (p_2) \label{Mb} \;, \\ 
M^{(c)} &=& \bar{u}(p_3) \slsh{\e}_1 \frac{\slsh{p_{13}}+m}{p_{13}^2-m^2} \slsh{\e}_4 v(p_2) \;,
\eeqn   
where $p_{13}=k_1+p_3$ and $p_{34}=p_3+k_4$.
From Eq.~(\ref{Massivespinordefs}) we write the massive spinors as,
\beqn
\bar{u}(p_3) &=& N_3 \bar{u}(k_2) (\slsh{p}_3+m) \;, \\
v(p_2) &=& N_2 (\slsh{p}_2-m) v(k_3) \;,
\eeqn
and then reorganize the calculation by using the identities~\cite{Morgan:1995te},
\begin{eqnarray}
(\slsh{p}_3+m) \slsh{\e}_4  
& \equiv & 
(\slsh{p}_{34}-\slsh{k}_4+m) \slsh{\e}_4 \nn \\
& \equiv & 
(2 p_{34}-k_4) \cdot \e_4 
 -\frac{1}{2} [\slsh{k}_4,\slsh{\e}_4]
 -\slsh{\e}_4  (\slsh{p}_{34}-m) \;, \\
(-\slsh{p}_{12}+m) \slsh{\e}_1  
& \equiv & 
-(2 p_{2}+k_{1}) \cdot \e_1 
 -\frac{1}{2} [\slsh{k}_1,\slsh{\e}_1]
 +\slsh{\e}_1  (\slsh{p}_{2}+m) \;.
\end{eqnarray}
Inserting these identities we may write new forms for $M^{(a)}$ and $M^{(c)}$,
\beqn \label{singleplace}
M^{(i)} &=& N_2 N_3 \bar{u}(k_2) \; \Gamma^{(i)} \; (\slsh{p}_2-m) \; v(k_3) \;,\;\;\; i=(a,b,c) \\
\Gamma^{(a)} &=& \Big\{ \Big[2 p_{3} \cdot \e_4  -\frac{1}{2} [\slsh{k}_4,\slsh{\e}_4]\Big]
 \frac{1}{2 p_3 \cdot k_4 } \Big[-2 p_{2} \cdot \e_1  -\frac{1}{2} [\slsh{k}_1,\slsh{\e}_1]\Big] 
-\slsh{\e}_4 \slsh{\e}_1 \Big\} 
\label{Ma} \;, \\
\Gamma^{(c)} &=& \Big\{ \Big[2 p_{3} \cdot \e_1  -\frac{1}{2} [\slsh{k}_1,\slsh{\e}_1]\Big]
 \frac{1}{2 p_3 \cdot k_1 } \Big[-2 p_{2} \cdot \e_4  -\frac{1}{2} [\slsh{k}_4,\slsh{\e}_4]\Big] 
-\slsh{\e}_1 \slsh{\e}_4 \Big\} \;. \label{Mc} 
\eeqn   
These formulae have the advantage that the $m$ dependence is corralled in a single place, 
in Eq.~(\ref{singleplace}).
Moreover the overall form is simple and the last term in Eqs.~(\ref{Ma}, \ref{Mc}) does 
not contain the massive propagator. $\Gamma^{(b)}$ can be read off from Eq.~(\ref{Mb}). 
Inserting the appropriate polarization vectors we find,
\begin{eqnarray}
\label{gQbQgresults}
-i A(1_g^-,2_\Qb^+,3_\Q^-,4_g^+) &=&
-\frac{\spb2.4^2 \spab1.\slsh{\bf 2}.4} {\spb1.4 \spb2.3 \spab4.\slsh{\bf 3}.4}
\;, \nn \\
-i A(1_g^-,2_\Qb^-,3_\Q^+,4_g^+) &=& 
-\frac{{\spa1.2}^2 \spab1.\slsh{\bf 3}.4} {\spa1.4 \spa2.3 \spab4.\slsh{\bf 3}.4}
\;, \nn \\
-i A(1_g^-,2_\Qb^-,3_\Q^-,4_g^+) &=&
  2 (\beta_+-\beta_-) \frac{m}{\spb2.3}
    \frac{{\spa1.3}^2 {\spb3.4}^2}{\spa1.4 \spb1.4 \spab4.\slsh{\bf 3}.4}
\;, \nn \\
-i A(1_g^-,2_\Qb^+,3_\Q^+,4_g^+) &=&
  2 (\beta_+-\beta_-) \frac{m}{\spa2.3}
          \frac{{\spa1.3}^2 {\spb3.4}^2}{\spa1.4  \spb1.4 \spab4.\slsh{\bf 3}.4}
\;, \nn \\
-i A(1_g^-,2_\Qb^+,3_\Q^-,4_g^-) &=& 0
\;, \nn \\
-i A(1_g^-,2_\Qb^-,3_\Q^+,4_g^-) &=& 0
\;, \nn \\
-i A(1_g^-,2_\Qb^+,3_\Q^+,4_g^-) &=& 
   -\frac{m \beta_+} {\spa2.3}\frac{{\spa1.4}^2}{\spab4.\slsh{\bf 3}.4}
\;, \nn \\
-i A(1_g^-,2_\Qb^-,3_\Q^-,4_g^-) &=& 
    \frac{m \beta_-} {\spb2.3}\frac{{\spa1.4}^2}{\spab4.\slsh{\bf 3}.4} \;.
\end{eqnarray}

\subsection{Results for $A(1_q,2_{\bar{q}},3_g)$}
The tree-level results for the simple $q\bar q g$ amplitudes, stripped of
overall colour and coupling constant factors, are well known:
\beqn
-i A(1_\q^-,2_\qb^+,3_g^-)&=& -\frac{{\spa1.3}^2}{\spa1.2} \;, \\
-i A(1_\q^-,2_\qb^+,3_g^+)&=& -\frac{{\spb2.3}^2}{\spb1.2} \;.
\eeqn 

\subsection{Results for $A(1_q,2_{\bar{q}},3_g,4_{\lb},5_l)$}
The tree-level amplitudes for $q\bar q W g$ are also rather simple. Removing
the colour and coupling constants as normal we have,
\beqn
-i A(1_\q^-, 2_\qb^+,3_g^-,4_\lb^+,5_\l^-) &=& 
-\frac{\spb2.4^2}{\spb1.3 \spb2.3 \spb4.5} \;, \nn \\
-i A(1_\q^-, 2_\qb^+,3_g^+,5_\lb^+,4_\l^-) &=& 
 \frac{\spa1.5^2}{\spa1.3 \spa2.3 \spa4.5} \;.
\eeqn

\section{Scalar integrals}
\label{intdef}
We work in the Bjorken-Drell metric so that
$l^2=l_0^2-l_1^2-l_2^2-l_3^2$. 
The definition of the integrals is as follows,
\begin{eqnarray}
&& A_0(m_1^2)  =
 \frac{\mu^{4-D}}{i \pi^{\frac{D}{2}}\rG}\int d^D l \;
 \frac{1}{(l^2-m_1^2)}\,, \nn \\
&& B_0(p_1;m_1,m_2)  =
 \frac{\mu^{4-D}}{i \pi^{\frac{D}{2}}\rG}\int d^D l \;
 \frac{1}
{(l^2-m_1^2)
((l+p_1)^2-m_2^2)}\,,\nn \\
&& C_0(p_1,p_2;m_1,m_2,m_3)  =
\frac{\mu^{4-D}}{i \pi^{\frac{D}{2}}\rG}
 \\
&& \times \int d^D l \;
 \frac{1}
{(l^2-m_1^2)
((l+p_1)^2-m_2^2)
((l+p_1+p_2)^2-m_3^2)}\,,\nn \\
&&\nn \\
&&
D_0(p_1,p_2,p_3;m_1,m_2,m_3,m_4)= 
\frac{\mu^{4-D}}{i \pi^{\frac{D}{2}}\rG}
\nn \\
&&
\times \int d^D l \;
 \frac{1}
{(l^2-m_1^2)
((l+p_1)^2-m_2^2)
((l+p_1+p_2)^2-m_3^2)
((l+p_1+p_2+p_3)^2-m_4^2)}\,. \nn
\end{eqnarray}

We have removed the overall constant which occurs in $D$-dimensional integrals 
\beq
\rG\equiv\frac{\Gamma^2(1-\e)\Gamma(1+\e)}{\Gamma(1-2\e)} = 
\frac{1}{\Gamma(1-\e)} +{\cal O}(\e^3) =
1-\e \gamma+\e^2\Big[\frac{\gamma^2}{2}-\frac{\pi^2}{12}\Big]
+{\cal O}(\e^3)\,.
\eeq
\section{Fermionic self energy}
\label{fermionse}
Introducing the renormalization of the bare parameters, $m_0= Z_m m$ and $Q_0=\sqrt{Z_Q} Q$
we may write the renormalized inverse propagator for a heavy quark of momentum $p$ as, 
\begin{equation}
-i \Gamma^R(p,m;g)=Z_Q \Big[ \slsh{p} -m - \Sigma(p,m;g) -m (Z_m-1)\Big] +O(g^4)\; .
\end{equation}
$-i \Sigma(p,m;g)$ is the contribution of the one-loop heavy quark
self-energy graph, which prior to renormalization has the form,
\begin{equation}
\Sigma(p,m;g)
 = -g^2 C_F \cg \Big[ X(p^2)  \, (\slsh{p}-m) + m \, Y(p^2) \Big] \;.
\end{equation}
By direct calculation of the Feynman diagram in an 
arbitrary covariant gauge specified by gauge fixing 
parameter $\lambda$,
\begin{eqnarray}
X(p^2)&=& \Bigg[2(1-\delta\e)(B_0(p;0,m)+B_1(p;0,m))-(1-\lambda) (B_0(p;0,m)+(p^2-m^2) B_0^\prime(p;0,m))\Bigg]
\nn \\
Y(p^2)&=&\Bigg[ 2B_1(p,0,m)(1-\delta\e)-2B_0(p;0,m)-(1-\lambda)(p^2-m^2) B_0^\prime(p;0,m))\Bigg]
\end{eqnarray} 
where $\delta=0$ in the FDH scheme, $\delta=1$
in the conventional dimensional regularization scheme, 
and $C_F=\frac{N_c^2-1}{2 N_c}$.
The integrals $B_0$ and $B_1$ are defined as,
\begin{equation}
\{B_0(p;0,m),B_1(p;0,m)p^\mu\} =
 \frac{\mu^{4-D}}{i \pi^{\frac{D}{2}}\rG}\int d^D l \;
 \frac{\{1,l^\mu\}}
{l^2 \; ((l+p)^2-m^2)} \;,
\end{equation}
and $B_i^\prime$ is the derivative of the form factor $B_i$ with respect 
to $p^2$.

Taking the limit $p^2=m^2$ before the limit $\e \to 0$, 
we have the following results,
\begin{eqnarray}
B_0(p;0,m)|_{p^2=m^2}&=& \Big( \frac{\mu^2}{m^2}\Big)\Big[ \frac{1}{\e} +2 \Big] \;, \nn\\
B_1(p;0,m)|_{p^2=m^2}&=& \Big( \frac{\mu^2}{m^2}\Big)\Big[ -\frac{1}{2 \e} -\frac{1}{2} \Big] \;, \nn\\
B_0^\prime(p;0,m)|_{p^2=m^2}&=& -\frac{1}{2 m^2}\Big( \frac{\mu^2}{m^2}\Big)\Big[ \frac{1}{\e} +2 \Big] \;, \nn \\
B_1^\prime(p;0,m)|_{p^2=m^2}&=& -\frac{1}{2 m^2} \;.
\end{eqnarray}
The mass renormalization is fixed by the condition that
the inverse propagator vanish on shell,
\begin{equation} \label{massrenormalization}
Z_m = 1+g^2 C_F Y(m^2)=1- \cg g^2 C_F
 \Bigg[ \frac{3}{\e} +3 \ln\left(\frac{\mu^2}{m^2}\right)+5-\delta\Bigg] +O(g^4,\e) \;.
\end{equation}
After mass renormalization the result for the inverse propagator becomes,
\begin{equation}
-i \Gamma^R(p,m;g) = Z_Q \Big[(\slsh{p}-m) (1 +g^2 C_F \cg X(p^2))  +m \, C_F \cg \big(Y(p^2)-Y(m^2)\big) \Big] \;.
\end{equation}
Renormalizing the wave function at the point $p^2=m^2$ we find,
\begin{equation}
Z_Q= 1-g^2 C_F \cg \Big[ X(m^2) + 2 m^2 \frac{d Y(p^2)}{d p^2}\Big|_{p^2=m^2}\Big] +O(g^4) \;.
\end{equation}
By explicit calculation we have that,
\begin{eqnarray}
X(m^2)&=& \Bigg[ \frac{1}{\e} +\ln\left(\frac{\mu^2}{m^2}\right)+ (3-\delta) -(1-\lambda)
\left(\frac{1}{\e} +\ln\left(\frac{\mu^2}{m^2}\right)+2\right)+O(\e) \Bigg]\;,  \nn \\
2 m^2 \frac{dY(p^2)}{dp^2} \Bigg|_{p^2=m^2}
&=& \Bigg[ \frac{2}{\e} + 2 \ln\left(\frac{\mu^2}{m^2}\right)+2
+(1-\lambda)
\left(\frac{1}{\e} +\ln\left(\frac{\mu^2}{m^2}\right)+2\right)
 +O(\e) \Bigg] \;.
\end{eqnarray}
The final result for the wave function renormalization is independent of the gauge fixing parameter, $\lambda$,
\begin{equation} \label{wavefunction}
Z_Q= 1-g^2 \cg C_F \Bigg[ \frac{3}{\e}+3 \ln\left(\frac{\mu^2}{m^2}\right)+5-\delta\Bigg]+O(g^4,\e)\; .
\end{equation}
The agreement between Eq.~(\ref{massrenormalization}) and Eq.~(\ref{wavefunction}) is fortuitous because 
the former contains only ultraviolet poles, whereas the latter contains both ultraviolet and infrared poles.
\section{Numerical evaluation}
\label{numer}
For the convenience of the reader we present numerical results for the amplitudes 
at a particular phase space point. 
The results we present contain no ultraviolet renormalization and 
the self energy corrections on the two massive external legs have not been included.
(The self energy corrections on the massless external legs vanish). 

The phase space point is specified by six momenta satisfying overall momentum conservation 
$k_1+p_2+p_3+k_4+k_5+k_6=0$ and $k_i^2=0$, $p_2^2=p_3^2=m^2$.
The massive momenta $p_2$ and $p_3$ are defined in Eqs.~(\ref{Rodrigodecomposition},\ref{betadef}).
For the numerical results the massless momenta $k_i$ are taken to be,
\begin{eqnarray}
\label{specificpoint}
     k_1 & = & \frac{\mu}{2} (-1,  +\sin\theta, +\cos\theta \sin\phi, +\cos\theta \cos\phi ), \nonumber \\
     k_2 & = & \frac{\mu}{3} (1,1,0,0), \nonumber \\
     k_3 & = & \frac{\mu}{7} (1,\cos\sigma,\sin\sigma,0), \nonumber \\
     k_4 & = & \frac{\mu}{2} (-1,  -\sin\theta, -\cos\theta \sin\phi, -\cos\theta \cos\phi ), \nonumber \\
     k_5 & = & \frac{\mu}{6} (1,\cos\rho \cos\sigma, \cos\rho \sin\sigma,\sin\rho), \nonumber \\
     k_6 & = & -k_1-k_2-k_3-k_4-k_5\, ,
\end{eqnarray}
where $\theta= \pi/4,\phi= \pi/6,\rho= \pi/3,\cos \sigma= -7/19, m=1$~GeV and $\mu = 6$~GeV.
Note that the energies of $k_1$ and $k_4$ are negative and $k_i^2=0$.
As usual $\mu$ also denotes the scale which is used
to carry the dimensionality of the $D$-dimensional integrals. 
With $m=1$~GeV we have $\beta \sim 0.38397382 \ldots$ so that $\beta$ differs substantially from the massless limit,
$\beta=1$.

The results for the primitive amplitudes are presented in Tables~\ref{numres1}
and~\ref{numres2}, where we have divided the 1-loop amplitudes by their
tree-level counterparts in order to remove the overall ambiguity in the phase~\footnote{Our phase for the 
spinor products can be understood from the routine~{\it spinoru.f}~~in the MCFM distribution.}.
The individual amplitudes labelled by the particular spin labels are dependent on our choice of
the spinor wave functions, Eq.~(\ref{Massivespinordefs}).
Results independent of this convention may be obtained by summing the squares of the four 
amplitudes.
 \begin{table}
 \begin{tabular}{|c||c|c|c||c|c|c|}
 \hline
 &\multicolumn{3}{c||}{$A_6(1_q^-,2_\Qb^+,3_Q^-,4_\qb^+,5_{\bar{\ell}}^+,6_{\ell}^-)$}
 &\multicolumn{3}{c||}{$A_6(1_q^-,2_\Qb^-,3_Q^+,4_\qb^+,5_{\bar{\ell}}^+,6_{\ell}^-)$}
 \\
 \hline
 \hline
 & $1/\e^2$ & $1/\e$ & $\e^0$
 &$1/\e^2$ & $1/\e$ & $\e^0$ \\
 \hline
 \hline $-i \Atree_6$ 
& $      0   $ & $      0   $ & $     -0.02873880764   $ & $      0   $ & $      0   $ & $      0.00703966555   $ \\ 
&  &  & $      0.00020478685\,i$ &  &  & $      0.08491754257\,i$ \\ 

 \hline $\Alc_6/ \Atree_6$
& $     -1   $ & $      3.84296991603   $ & $     21.70930260416   $ & $     -1   $ & $      3.84296991603   $ & $     21.65220761331   $ \\ 
&  &  & $      6.25248592358\,i$ &  &  & $      2.82472065470\,i$ \\ 

 \hline $\Asl_6/ \Atree_6$
& $     -1   $ & $      0.72362882832   $ & $     22.03005675285   $ & $     -1   $ & $      0.72362882832   $ & $     21.99641271870   $ \\ 
&  & $     -7.83563162387\,i$ & $    -22.49326728353\,i$ &  & $     -7.83563162387\,i$ & $    -22.71671278884\,i$ \\ 

 \hline $\Alf_6/ \Atree_6$ 
& $      0   $ & $      0.66666666667   $ & $      2.12677656298   $ & $      0   $ & $      0.66666666667   $ & $      2.12677656298   $ \\ 
&  &  & $      2.09439510239\,i$ &  &  & $      2.09439510239\,i$ \\ 

 \hline
 \hline
 \end{tabular}
 \caption{Numerical values of primitive amplitudes
 at the kinematic point defined in
 Eqs.~(\ref{Rodrigodecomposition},\ref{specificpoint}).
 \label{numres1}}
 \end{table}
 \begin{table}
 \begin{tabular}{|c||c|c|c||c|c|c|}
 \hline
 &\multicolumn{3}{c||}{$A_6(1_q^-,2_\Qb^-,3_Q^-,4_\qb^+,5_{\bar{\ell}}^+,6_{\ell}^-)$}
 &\multicolumn{3}{c||}{$A_6(1_q^-,2_\Qb^+,3_Q^+,4_\qb^+,5_{\bar{\ell}}^+,6_{\ell}^-)$}
 \\
 \hline
 \hline
 & $1/\e^2$ & $1/\e$ & $\e^0$
 &$1/\e^2$ & $1/\e$ & $\e^0$ \\
 \hline
 \hline $-i \Atree_6$ 
& $      0   $ & $      0   $ & $     -0.03716837450   $ & $      0   $ & $      0   $ & $     -0.03716837450   $ \\ 
&  &  & $      0.03263223728\,i$ &  &  & $      0.03263223728\,i$ \\ 

 \hline $\Alc_6/ \Atree_6$
& $     -1   $ & $      3.84296991603   $ & $     16.95522192600   $ & $     -1   $ & $      3.84296991603   $ & $     24.66977363941   $ \\ 
&  &  & $      5.96305954078\,i$ &  &  & $      4.27788837815\,i$ \\ 

 \hline $\Asl_6/ \Atree_6$
& $     -1   $ & $      0.72362882832   $ & $     21.88454128871   $ & $     -1   $ & $      0.72362882832   $ & $     21.88454128871   $ \\ 
&  & $     -7.83563162387\,i$ & $    -21.57825725817\,i$ &  & $     -7.83563162387\,i$ & $    -21.57825725817\,i$ \\ 

 \hline $\Alf_6/ \Atree_6$ 
& $      0   $ & $      0.66666666667   $ & $      2.12677656298   $ & $      0   $ & $      0.66666666667   $ & $      2.12677656298   $ \\ 
&  &  & $      2.09439510239\,i$ &  &  & $      2.09439510239\,i$ \\ 

 \hline
 \hline
 \end{tabular}
 \caption{Numerical values of primitive amplitudes
 at the kinematic point defined in
 Eqs.~(\ref{Rodrigodecomposition},\ref{specificpoint}).
 \label{numres2}}
 \end{table}
The results in Tables~\ref{numres1} and \ref{numres2} have been checked by an independent program.

\end{document}